\documentclass[12pt]{iopart}
\usepackage{epsfig}

\newcommand{\be}{\begin{equation}}
\newcommand{\ee}{\end{equation}}
\newcommand{\bea}{\begin{eqnarray}}
\newcommand{\eea}{\end{eqnarray}}

\newcommand{\beq}{\begin{equation}}
\newcommand{\eeq}{\end{equation}}

\newcommand{\nn}{\nonumber}

\def\srho{\sqrt{s(s-4m^2)}}

\def\la{\mathrel{\mathpalette\fun <}}

\def\fun#1#2{\lower3.6pt\vbox{\baselineskip0pt\lineskip.9pt
\ialign{$\mathsurround=0pt#1\hfil##\hfil$\crcr#2\crcr\sim\crcr}}}

\begin{document}

\title[The $\rho\to\gamma\pi$ and $\omega\to\gamma\pi$ decays  in 
quark-model approach]{The $\rho\to\gamma\pi$ and $\omega\to\gamma\pi$ 
decays  in quark-model approach and estimation of coupling for pion 
emission by quark}

\author{A V  Anisovich, V V  Anisovich, L G  Dakhno,   M A  Matveev,\\ 
V A  Nikonov and A V  Sarantsev}

\address{Petersburg Nuclear Physics Institute, 188300, Gatchina, Russia}

\begin{abstract}
In the framework of the relativistic and gauge invariant spectral
integral technique, we calculate radiative decays
 $\rho (770)\to \gamma \, \pi (140)$ and $\omega (780)\to \gamma \, \pi (140)$
 supposing all mesons ($\pi$, $\rho$ and $\omega$) to be quark--antiquark
 states. The $q\bar q$ wave functions found for mesons and photon
lead to a reasonably good description of data
($\Gamma^{(exp)}_{\rho^{\pm} \to\gamma\pi^{\pm}}=68\pm 30 $ keV,
$\Gamma^{(exp)}_{\rho^{0} \to\gamma\pi^0}=77\pm 28  $ keV,
$\Gamma^{(exp)}_{\omega \to\gamma\pi^0}=776\pm 45 $ keV) that makes it
possible to estimate the coupling for the bremsstrahlung
emission of pion by quarks  $g_\pi\equiv g_\pi (u\to d \pi )$. We have found
two values for the pion bremsstrahlung
coupling: $|g_\pi|=16.7 \pm 0.3\ ^{+0.1}_{-2.3}$ (Solution I)  and
$|g_\pi|=3.0  \pm 0.3\ ^{+0.1}_{-2.1}$ (Solution II). Within
SU(6)-symmetry for nucleons, Solution I gives us for $\pi NN$ coupling
the value $16.4 \le  g_{\pi NN}^2/(4\pi) \le 23.2$ that is in qualitative
agreement with the $\pi N$ scattering data,
$g_{\pi NN}^2/(4\pi)\simeq 14$.
For excited states, we have estimated
the partial widths in Solution I as follows:
$\Gamma (\rho_{2S}^\pm\to \gamma\pi)\simeq 10   - 130$ keV,
$\Gamma (\rho_{2S}^0\to \gamma\pi)\simeq 10   -130$ keV,
$\Gamma (\omega_{2S}\to \gamma\pi)\simeq 60 - 1080$ keV.
The large uncertainties emphasise the
necessity to carry out
measurements of the meson radiative processes in the region of large masses.
 \end{abstract}

\submitto{\JPG}
\maketitle

\section{Introduction}

The radiative decay amplitude is a necessary element
for the study of the quark--gluon structure of hadrons.
In this paper, we  present the
calculation of the  radiative decays of
quark--antiquark states $(q\bar q)_{in}=\rho,\omega$
into $\gamma\pi$. In this way, we continue the calculations initiated in
\cite{raddecay}
where  radiative transitions of quarkonium states 
$(Q\bar Q)_{in}\to\gamma (Q\bar Q)_{out}$  were studied,
with the production of massive
outgoing states $(Q\bar Q)_{out}$. Considering the production of the
$\gamma\pi$ system, a particular necessity is to
take into account, together with the annihilation $q\bar q \to\pi$, an
additional process of the  bremsstrahlung type, namely, $q\to q \pi  $.

We treat the meson decay amplitude as triangle
diagram of constituent quarks  (additive quark model) calculated
in terms of the spectral integration technique, see \cite{book3}
and references therein.
The spectral integral technique is rather profitable for the
description of composite particles, for the
content of a composite system is thus strictly controlled. Besides,
this technique is rather convenient for the description of high spin
states.

The equation for the composite $q\bar q$ systems in
the spectral integration technique was suggested in \cite{BS}, it is a
direct generalisation  of the dispersion $N/D$ equation \cite{CM} when
 the $N$-function was represented as an infinite sum of separable
 vertices,
see \cite{book3} for detail. In terms
of this equation, the $b\bar b$ and $c\bar c$ quarkonia were
considered in \cite{QQ}, while the light-quark $q\bar q$ mesons were
studied in \cite{qq}.

In \cite{qq}, the levels of the one-component  $q\bar q$
systems (with $I=1$ or $I=0$
which are almost pure $s\bar s$ or $n\bar n=(u\bar u+d\bar d)/\sqrt 2$ states)
were reconstructed as well as their wave functions. The
$q\bar q$ systems are formed at distances, where perturbative QCD does
not work ($r\sim 0.5-1.0$ fm). In this region (the region of soft
interactions), we deal with constituent quarks and  effective
massive gluons (with mass of the order of 700--1000 MeV
\cite{h-e,gluon,cornwell,gerasyuta,g-lat}). It means that
quark--antiquark interactions undergone a significant changes as
compared to small distances; besides, at large distances the
confinement forces work. Therefore, interactions in the soft region
should be reconstructed on the basis of experimental studies -- in
\cite{qq}, the  $q\bar q$ interaction was reconstructed  on the basis
of available data for $q\bar q$-levels and the $q\bar q$-meson
radiative decays.

The standard way to investigate quark--antiquark systems is to
apply the Bethe-Salpeter equation \cite{bethe} written in terms of
Feynman integrals. One may find the examples of such a study of light
quark--antiquark systems in \cite{Hulth,Godfrey,Lucha,petry-meson,G}
and for heavy quarkonia ($c\bar c$ and $b\bar b$) in
\cite{G,Linde,M,Gupta,Sch,Huang}, see also references therein.

However, one should keep in mind an important difference between the
standard Bethe--Salpeter equation
and that written in terms of the spectral integral \cite{BS}.
In the dispersion relation technique, the constituents
in the intermediate state are mass-on-shell, $k^2_i=m^2$, while in the
Feynman technique, which is used in the Bethe--Salpeter equation,
$k^2_i\ne m^2$. So, in the spectral integral equation, when the high
spin state structures are calculated, we have a numerical factor
$k^2_i= m^2$, while in the Feynman technique one should write
$k^2_i= m^2+(k_i^2-m^2)$. Here, the first term in the right-hand side
provides us the contribution similar to
that used in the spectral integration technique, while the second
term cancels one of denominators of the kernel of the Bethe--Salpeter
equation, that results in the penguin or tadpole type diagrams --
let us call them zoo-diagrams. A particular property of the spectral
integral technique is the exclusion of  zoo-diagrams from the
equation for composite systems.

The spectral integral equation \cite{BS} gives us a unique
solution for the quark--antiquark levels and their wave functions,
provided the interquark interaction is known. Let us emphasize
that the equation works for both instantaneous interactions
and the  $t$-channel exchanges with retardation,  and even
for the energy-dependent interactions: this  follows from the fact
that the equation itself is the modified dispersion
relation for the amplitude. For solving the inverse problem, that
is, for  reconstructing the interaction, it is not enough to know the
meson masses --- one should know wave functions of quark--antiquark
systems. Such an information is contained  in the hadronic form
factors and radiative decay amplitudes. Therefore, in the
approach of refs. \cite{BS,QQ,qq}, we consider simultaneously the meson
levels in terms of the spectral integral equations and the meson 
radiative transitions in terms of the double dispersion relations over 
$q\bar q$ states (or over corresponding meson masses) --- in this 
way all calculations are carried out within compatible methods.

The calculation of  radiative transition amplitudes in
terms of the double dispersive integrals was performed for some
selected reactions in \cite{deut,physrev,epja,YFscalar,YFtensor} ---
the basic points of the method of operator expansion used in
the calculation of double dispersive integrals can be found in
\cite{raddecay,book3,operator}.

The analyses of the light $q\bar q$ systems \cite{qq} and heavy
$Q\bar Q$ quarkonia \cite{QQ} in terms of the spectral integral
equation differ from one another in certain respect, because the
available experimental data are of different sort: for the $Q\bar Q$
systems the only known are low-lying states
(with an exception for the $1^{--}$ quarkonia $\Upsilon$ and $\psi$
where a long series of vector states was discovered in the $e^+e^-$
annihilation). At the same time, for the low-lying states there exists
a rich set of data on radiative decays:
$(Q\bar Q)_{\rm in}\to\gamma (Q\bar Q)_{\rm out}$ and
$(Q\bar Q)_{\rm in}\to\gamma\gamma$. For the
light quark sector ($q\bar q$ systems), there exists an abundant
information on masses of highly excited states with different $J^{PC}$
(see \cite{PNPI-RAL,Bar,LL,L3-KK,L3-3pi} and surveys
\cite{book3,bugg,klempt-z}),
but the knowledge of radiative decays is rather poor.

Despite the scarcity of data on radiative decays,
the light $q\bar q$ states have been studied in \cite{qq},  relying upon
our knowledge of linear trajectories in the  $(n,M^2)$-plane, where
$n$ is the radial quantum number of the $q\bar q$-meson with
mass $M$ (see \cite{book3,syst}). We hope that it may somehow
compensate the lack of information on the wave functions.
In the fitting procedure \cite{qq}, the main attention was paid
to the states with large masses, expecting to extract the
confinement interaction. We obtained that the strong $t$-channel
interaction (which, as we think, determines the confinement) should 
exist in both scalar $I\otimes I$ and vector 
$\gamma_\mu\otimes  \gamma_\mu$ channels. The fitting results point 
rather reliably to the 
equality of these $t$-channel interactions \cite{qq}.

\begin{figure}[h]
%Fig.1
\centerline{\epsfig{file=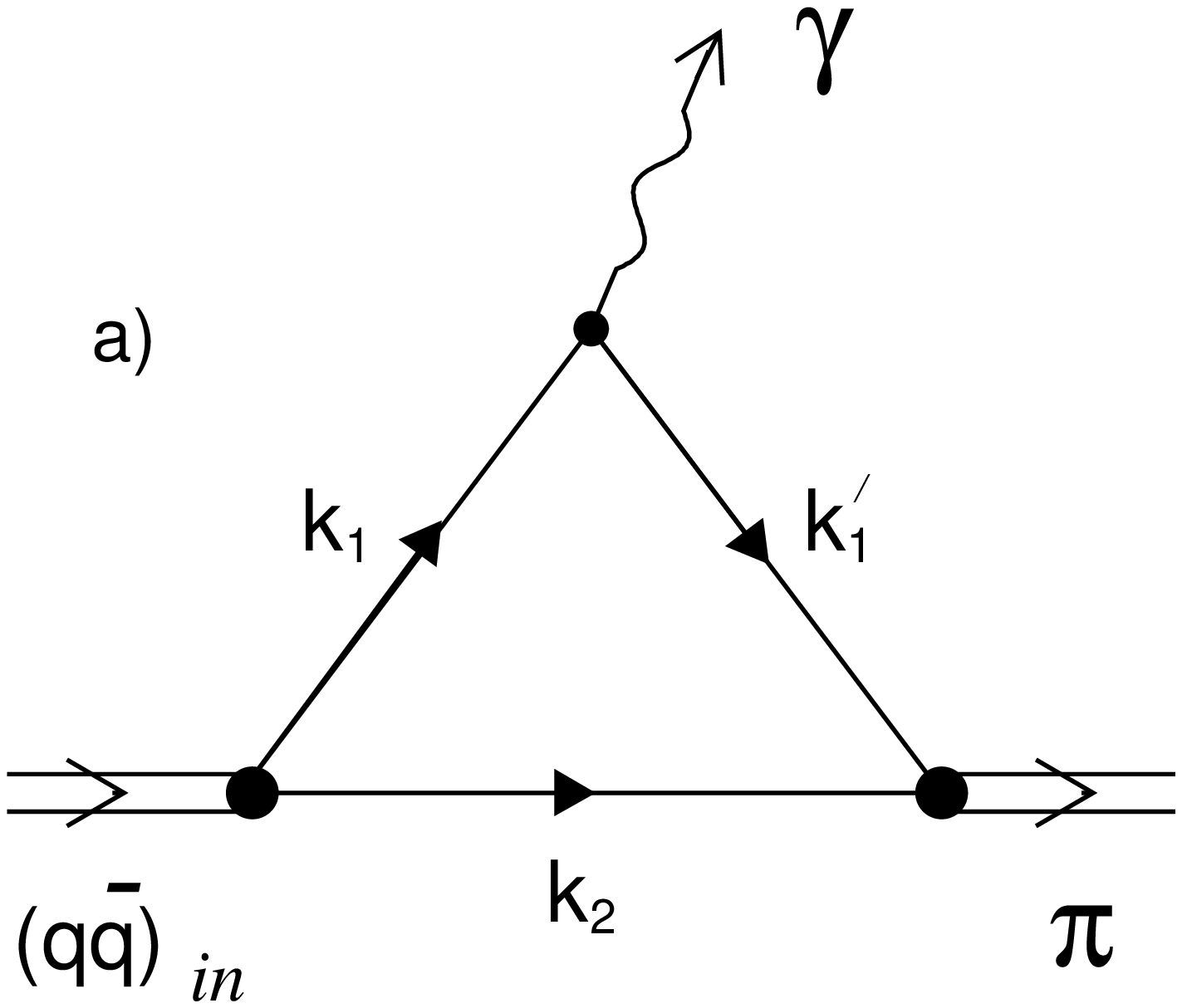,width=6cm}
            \epsfig{file=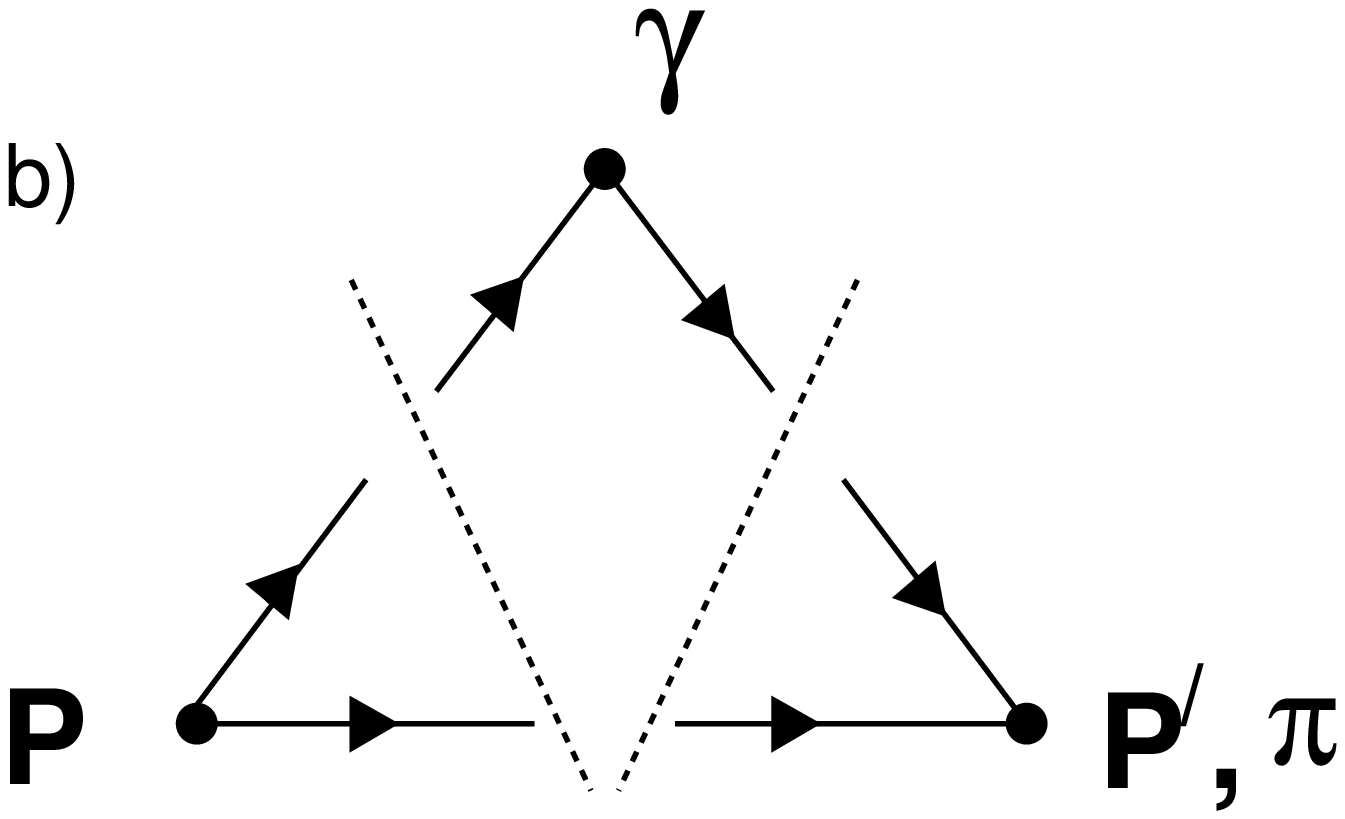,width=6cm}}
\centerline{\epsfig{file=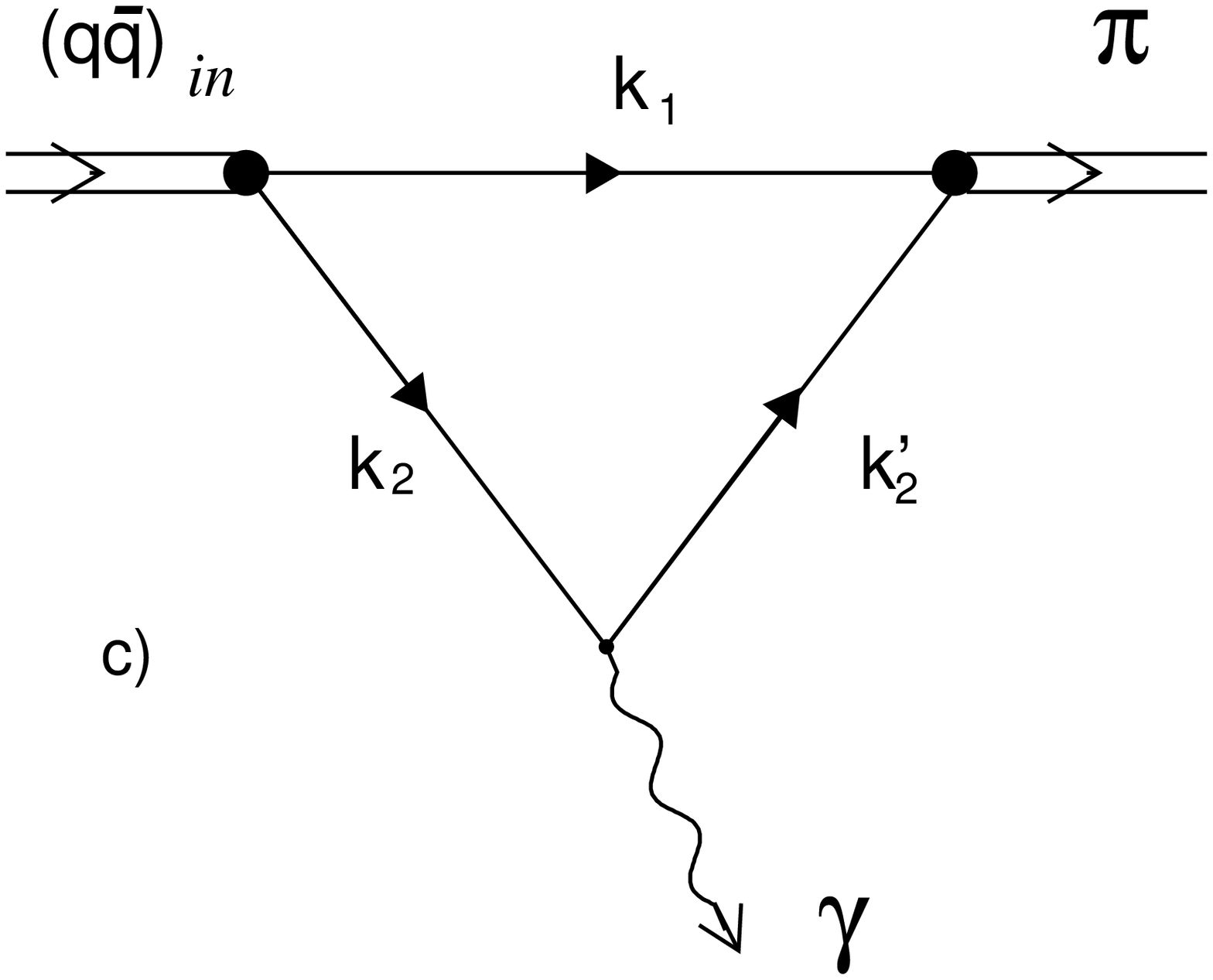,width=6cm}
            \epsfig{file=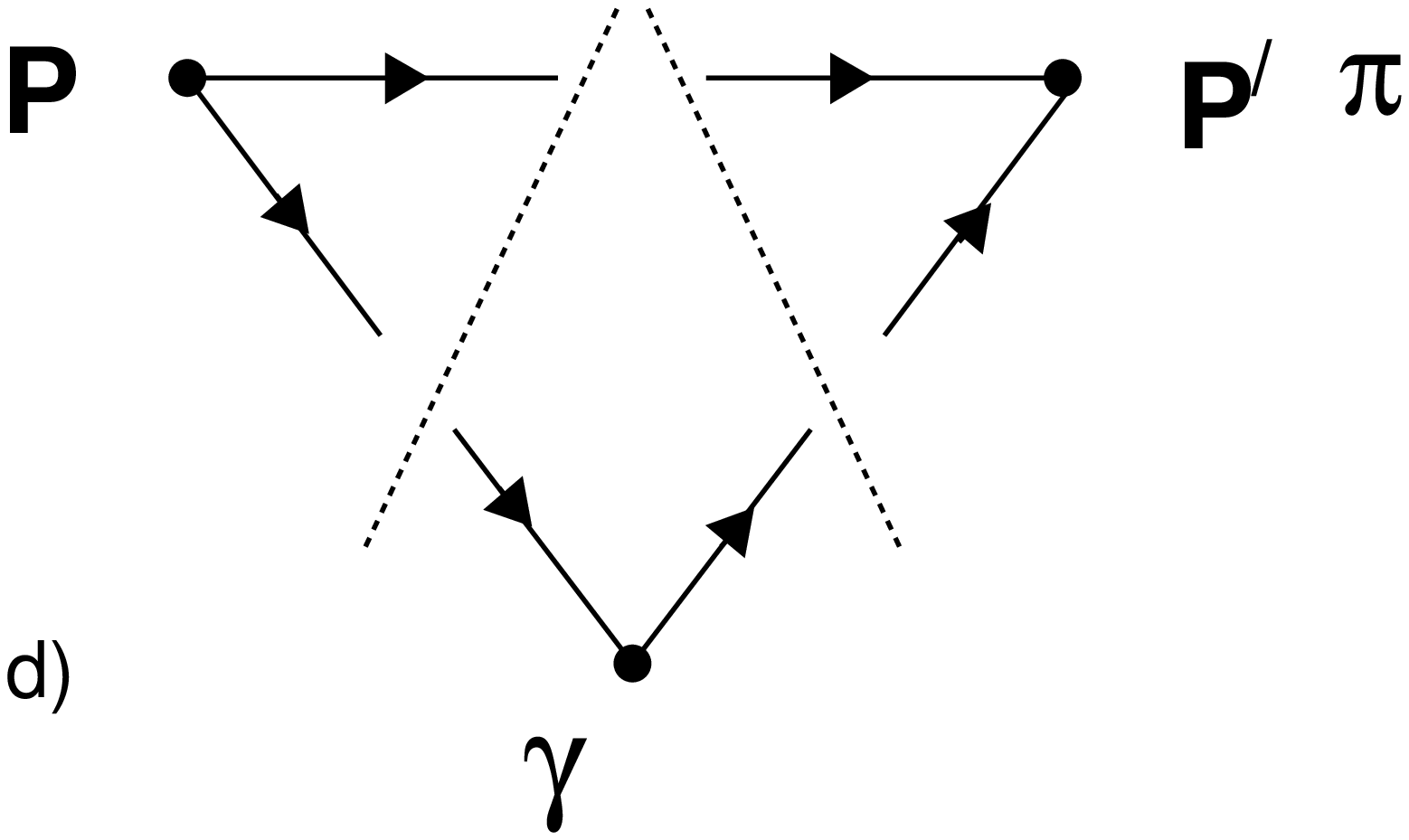,width=6cm}}
\caption{a), c) Triangle diagrams for radiative transition
$(q\bar q)_{in}\to \gamma \pi$ with the emission of photon by quark;
here $p^2=M^2_{in}$, $p'^2=M^2_{\pi}$ and $(p-p')^2=q^2$.
b), d) Cuttings of the triangle diagrams \ref{1}a and \ref{1}c signify
 the double discontinuity of
the spectral integral with intermediate-state momentum squared $P^2=s$,
$P'^2=s'$ and $(P-P')^2=q^2$.}
\label{1}
\end{figure}

\begin{figure}[h]
%Fig.2
\centerline{\epsfig{file=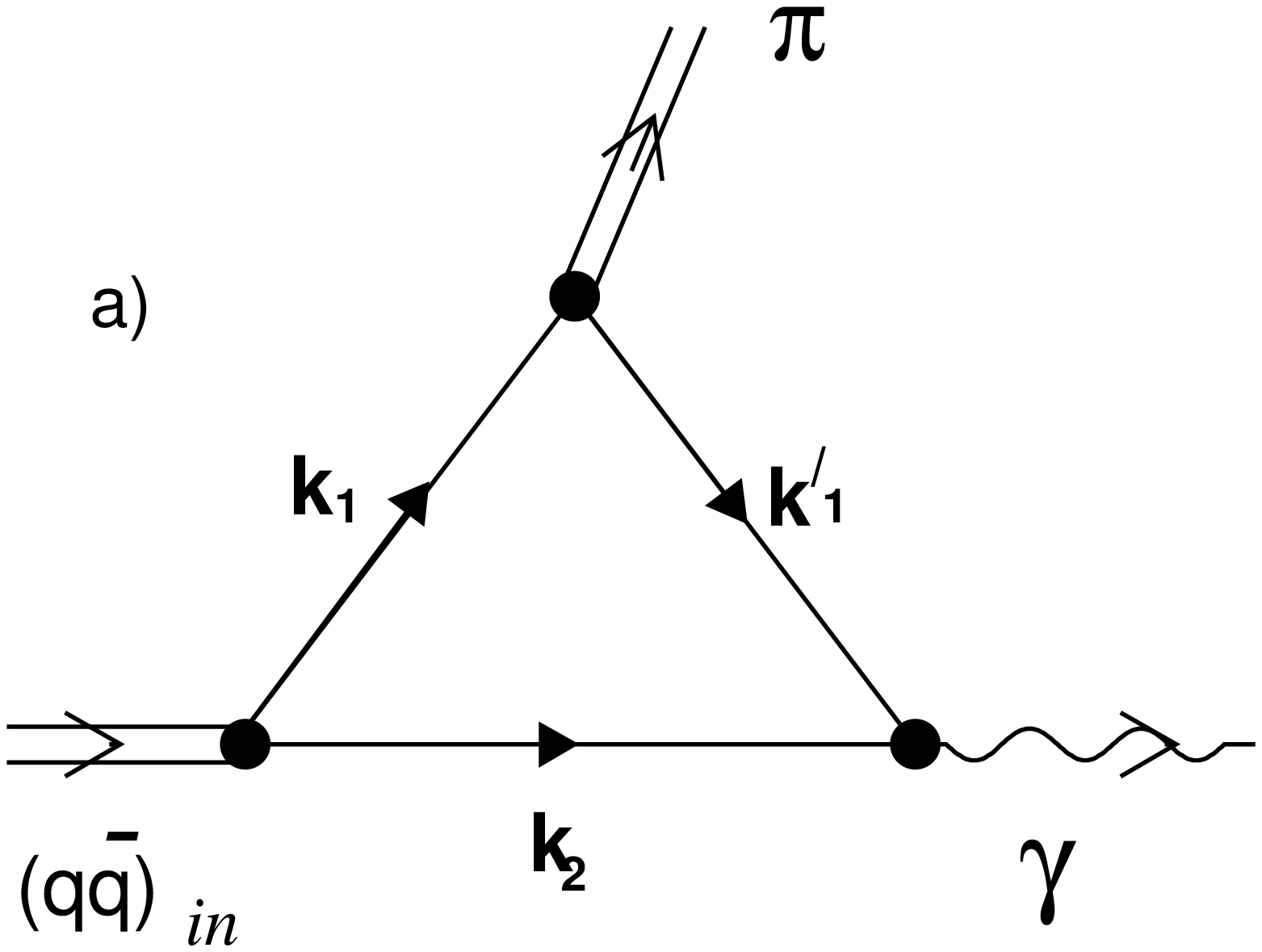,width=6cm}
            \epsfig{file=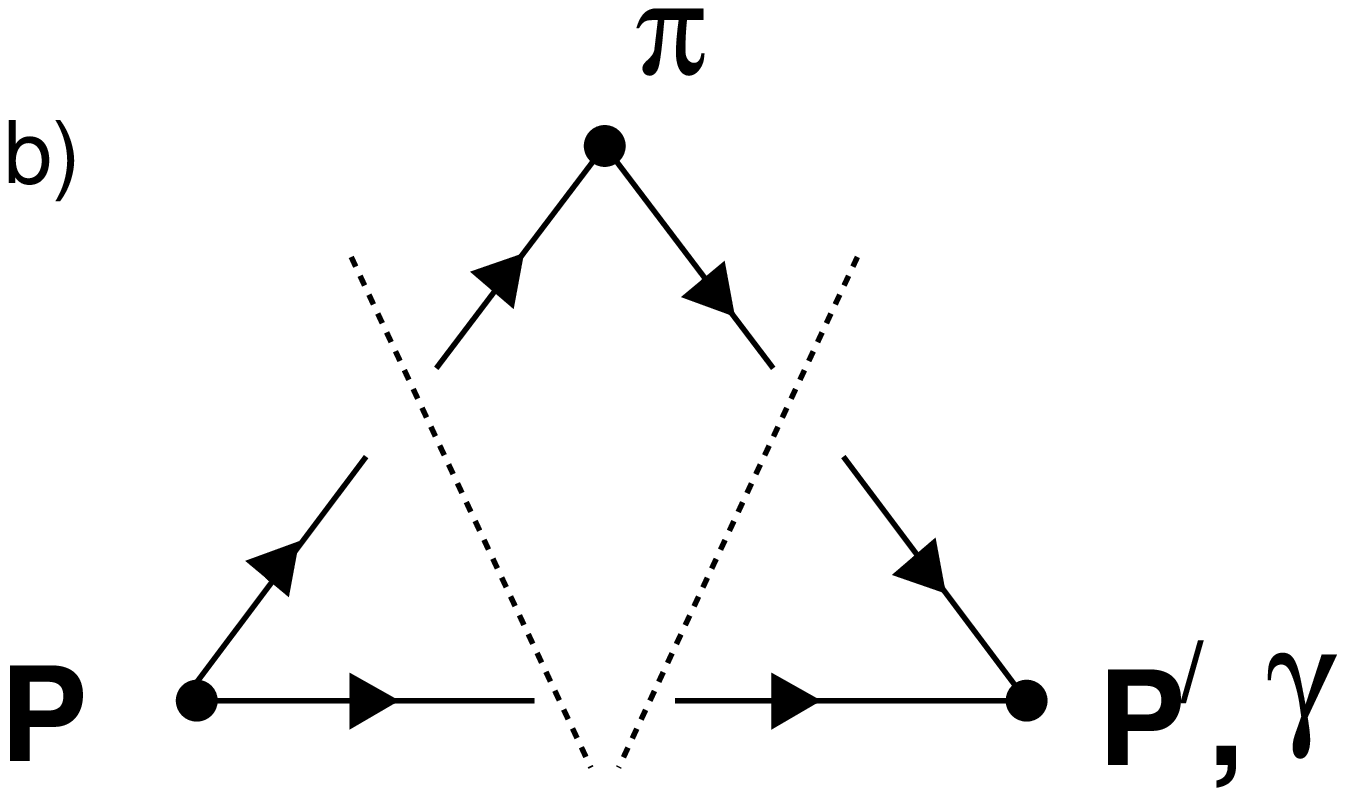,width=6cm}}
\centerline{\epsfig{file=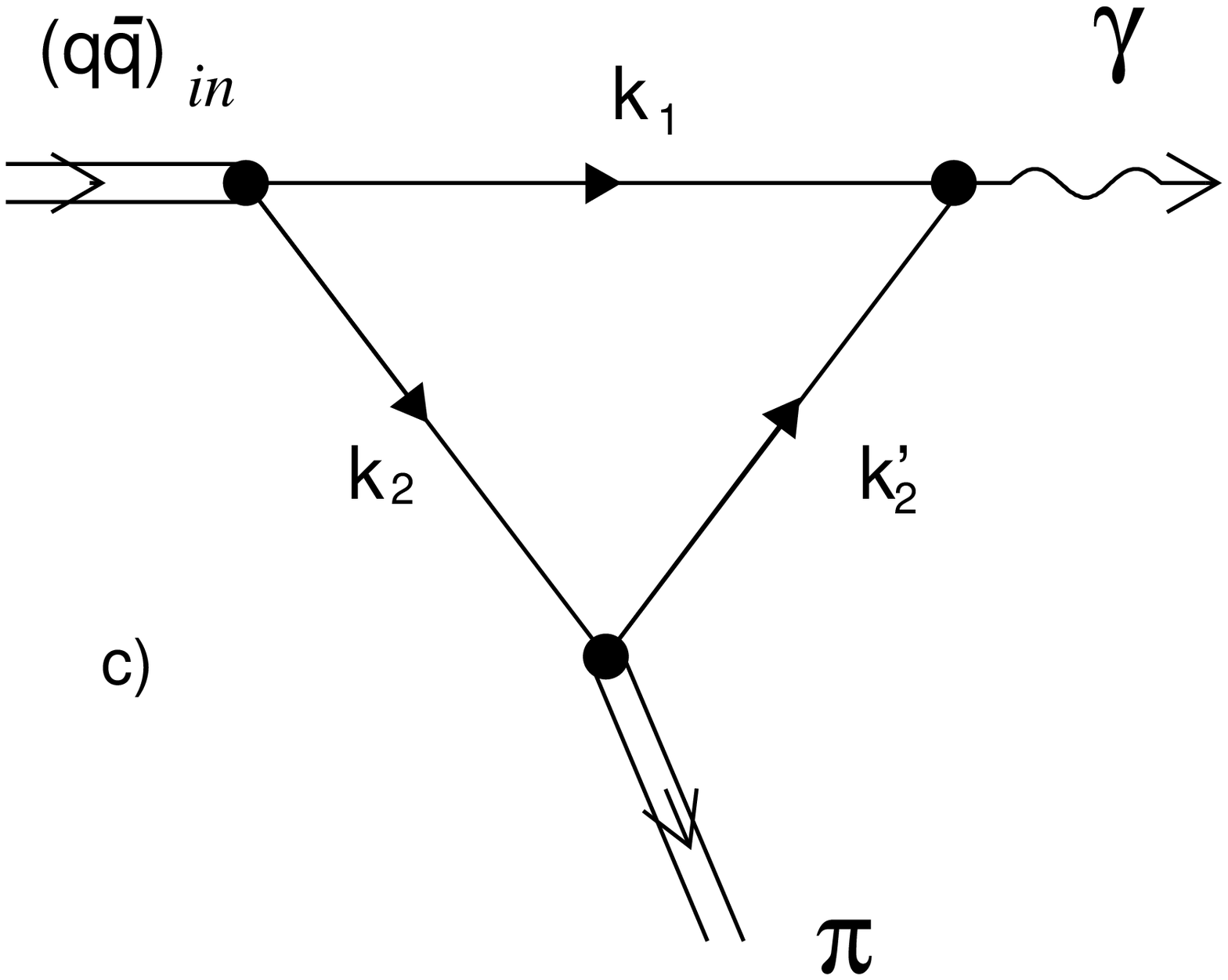,width=6cm}
            \epsfig{file=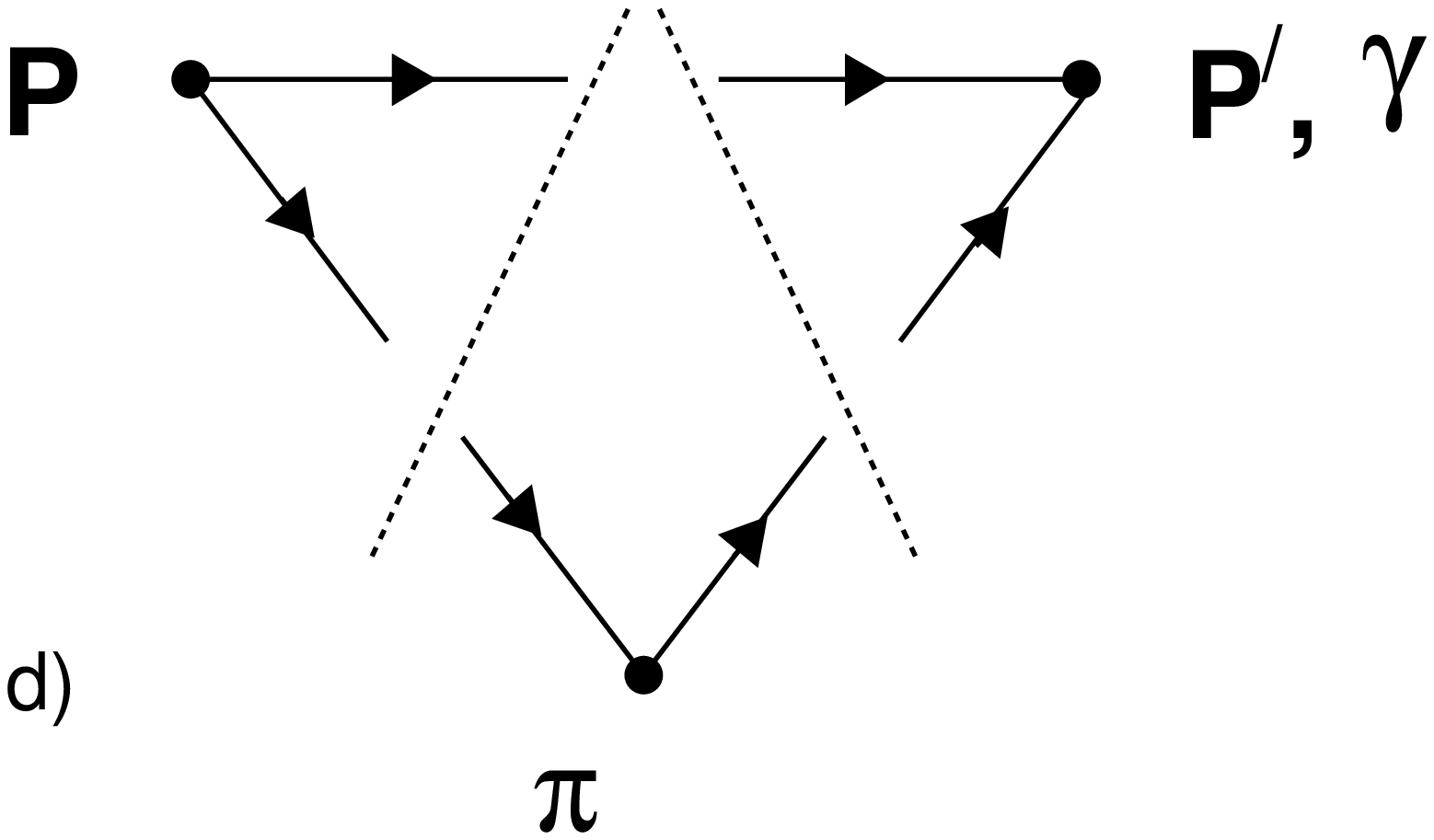,width=6cm}}
\caption{a), c) Triangle diagrams for  radiative transition
$(q\bar q)_{in}\to \gamma \pi$ with the emission of pion by quark;
here $p^2=M^2_{in}$, $p^2_\pi=M^2_\pi$.
b), d) Cuttings of the triangle diagrams \ref{2}a and \ref{2}c for getting
double discontinuity of the spectral integral with $P^2=s$, $P'^2=s'$
and $(P-P')^2=p^2_\pi$.} \label{2} \end{figure}

Obviously, the fitting results presented in \cite{qq} should be
checked (and, if necessary, improved) by investigating the other
radiative decays -- following to this program we consider here the
decays $\rho\to \gamma\pi$ and $\omega\to \gamma\pi$. Small mass of the
pion requires to take into account not only the process of photon
emission with a subsequent quark--antiquark annihilation
$q\bar q\to \pi$ (triangle diagram of the additive quark model,
Fig. \ref{1}) but
also the bremsstrahlung-type emission of pion $q\to \pi q$, with
subsequent quark--antiquark annihilation into photon
$q\bar q\to \gamma$, see Fig. \ref{2}.  Therefore, the key points in
the calculation of the $\rho,\omega\to \gamma\pi$ decays is to know
$q\bar q$ wave functions of pion and vector mesons ($\rho$ and
$\omega$) as well as $q\bar q $ wave function of the  photon
$\gamma\to q\bar q$. Also the fitting procedure calls us  to
determine the pion bremsstrahlung constant for the process $q\to \pi
q$.

\subsection{ Photon wave function}

For
the region $0\la Q^2 \la 1$ (GeV/{\it c})$^2$ (here $Q^2=-q^2$),
the light-quark components of
the photon wave function $\gamma^*(Q^2)\to q\bar q$  ($q=u,d,s$)
are determined in \cite{g-qq} (see also \cite{book3})
 on the basis of data for the transitions
 $\pi^0, \eta,\eta'\to \gamma\gamma^*(Q^2)$
and  reactions of $ e^+e^-$-annihilation:
$ e^+e^-\to \rho^0,\omega,\phi$ and $ e^+e^-\to hadrons$ at
$1<E_{e^+e^-}<3.7$ GeV (in a more rough approximation the wave function
$\gamma(Q^2)\to q\bar q$ was found in \cite{PR}).

 Conventionally, one may
consider two pieces of the photon wave function: soft and hard ones.
Hard component relates to the point-like vertex $\gamma\to q\bar q$, it
is responsible for the production of quark--antiquark pair
at high virtuality.
At high energies of the $e^+e^-$ system, the ratio of cross
sections $R=\sigma(e^+e^-\to hadrons)/\sigma(e^+e^-\to \mu^+\mu^-)$ is
determined by the hard component of photon wave function, while soft
component is responsible for the production of low-energy
quark--antiquark vector states such as
 $\rho^0$, $\omega$, $\phi(1020)$, and their  excitations.

In the spectral integral technique, the
quark wave function of the
photon, $\gamma^*(Q^2)\to q\bar q$, is defined as follows:
\be
\label{I2}
\psi_{\gamma^*(Q^2)\to q\bar q}(s)=
\frac{G_{\gamma\to q\bar q}(s)}{s+Q^2}\ ,
\ee
where $G_{\gamma \to q\bar q}(s)$ is the vertex
for the transition of photon into
$q\bar q$ state, depending on the
invariant energy squared, $s$, of $q\bar q$ system.
In terms of the light-cone variables $s=(m^2+k_\perp^2)/[x(1-x)]$,
where $m$ is the quark mass, ${k}_\perp$ and $x$  are the
light-cone characteristics of quarks: transverse
momentum and a part of longitudinal momentum.

Rather schematically, the vertex function
$G_{\gamma\to q\bar q}(s)$ may be divided into two terms.
The first term is responsible for the soft component
which is due to the transition of photon to vector $q\bar q$ meson
$\gamma\to V\to q\bar q$, while the second one describes
the point-like interaction in the hard domain.
The principal characteristics of the
soft component
is the threshold value of the vertex
and the rate of its decrease with
energy.
The hard component of the vertex is characterized by the
energy where  the point-like interaction becomes dominant.

In \cite{PR}, the photon wave function has been found assuming
the quark relative momentum dependence to be the same for all quark
vertices: $g_{\gamma\to u\bar u}(k^2)=g_{\gamma\to d\bar d}(k^2)$
$=$ $g_{\gamma\to s\bar s}(k^2)$, where we redenoted
$G_{\gamma\to q\bar q}(s)\longrightarrow g_{\gamma\to q\bar q}(k^2)$
with $k^2=s/4-m^2$.
The hypothesis of the vertex
universality for $u$ and $d$ quarks used in \cite{g-qq},
\be
G_{\gamma\to u\bar u}(s)=G_{\gamma\to d\bar d}(s)\equiv
G_{\gamma}(s)\ ,
\ee
looks rather trustworthy because
of the degeneracy of $\rho$ and $\omega$
states, though the similarity in the $k$-dependence  for  non-strange
and strange quarks may be violated. Using experimental
data on the transitions $\gamma\gamma^*(Q^2)\to \pi^0,\eta,\eta'$ only,
one cannot determine the parameters  ($C,b,s_0$ -- see below Eqs. 
(\ref{51}) and (\ref{52})) for both
$G_{\gamma\to s\bar s}(s)$ and $G_{\gamma}(s)$.
We also add the $e^+e^-$
annihilation data for the determination of wave functions, that is
$e^+e^-\to \gamma^*\to \rho^0,\omega,\phi(1020)$, together with the
ratio
$R(E_{e^+e^-})=\sigma(e^+e^-\to hadrons)/\sigma(e^+e^-\to\mu^+\mu^-)$ at
$E_{e^+e^-}> 1$ GeV. The reactions
$e^+e^-\to \gamma^*\to \rho^0,\omega,\phi(1020)$ are rather sensitive
to the parameters $C_a,b_a$,
 while the data on $R(E_{e^+e^-})$ allow us to fix the
parameter $s_0$.

The transition vertices for $u\bar u,d\bar d\to\gamma$
have been chosen in the form:
\bea
u\bar u,\, d\bar d : \qquad
G_\gamma(s)&=&c_\gamma \bigg(\exp({-b^\gamma_1 s})+c^\gamma_2
\exp({-b^\gamma_2 s})\bigg)+\frac1{1+e^{-b^\gamma_0(s-s_0^\gamma)}}\ ,
\label{51}
\eea
and the following
 parameter values  have been found \cite{book3,g-qq}:
\bea
\label{52}
u\bar u,\, d\bar d: \,\,         && c^\gamma=32.506,\;
c^\gamma_2=-0.0187,\;b^\gamma_1=4\, {\rm GeV}^{-2},\;
b^\gamma_2=0.8\, {\rm GeV}^{-2},\;   \\ \nonumber
&& b^\gamma_0=15\,{\rm GeV}^{-2},\; s_0^\gamma=1.614\,{\rm GeV}^{2}\ .
\eea
With these parameters,
 we have
a good description of the available
experimental data for $V\to e^+e^-$ and two-photon decays, see \cite{book3,qq}.

\subsection{ The $\rho$, $\omega$ and $\pi$ wave functions}

We characterise $q\bar q$-states by the following momentum-dependent
wave functions:
\be \label{Rwf}
\psi^{(S,L,J)}_n (k^2) =
\frac{G^{(S,L,J)}_n (k^2)}{s-\left(M_n^{(S,L,J)}\right)^2}\ ,
\ee
 where $S$, $L$, $J$
are the spin, orbital momentum and total momentum of the $q\bar q$ system with mass
$M^{(S,L,J)}_n$.

\subsubsection{$\rho(nL)$ and $\omega(nL)$ states}
We introduce
spin-orbital operators and wave functions for the states with dominant
$L=0,2$ as follows:
\bea \label{Rwf-1}
\begin{array}{l|l|r}\qquad L=0  &  0^{-+}  &
i\gamma_5\psi^{(0,0,0)}_n (k^2)  \\
{\rm dominant}\, L=0  &  1^{--}  &
 \gamma_\mu^\perp\psi^{(1,0,1)}_n (k^2)   \\
                    \hline
{\rm dominant}\, L=2     &  1^{--}  &
3/\sqrt{ 2}\cdot \left (k^\perp_\mu\hat k^\perp-
\frac13 k_\perp^2\gamma^{\perp}_\mu\right )\psi^{(1,2,1)}_n (k^2).
\end{array}
\eea
Here $k^\perp$ is the relative quark--antiquark momentum,
$k^\perp_\mu =( g_{\mu\mu'}-P_\mu P_{\mu'}/P^2)k_{1\mu'}
\equiv g_{\mu\mu'}^\perp k_{1\mu'} = -g_{\mu\mu'}^\perp k_{2\mu'} $,
so $k^\perp \perp P=k_1+k_2$; likewise,
$\gamma_\mu ^\perp =  g_{\mu\mu'}^\perp \gamma_{\mu'} $.
Definition of spin--momentum operators for other states can be found  in
\cite{book3,operator}.

Generally, the states with different $L$ mix with each other:
\bea \label{Rwf-1mix}
\hat \psi^{V(n,1)}_{\mu}(s)=C_{10}^{(n)} \gamma_\mu^\perp\psi^{(1,0,1)}_n (k^2)+
C_{12}^{(n)}
\frac{3}{\sqrt{ 2}} \left (k^\perp_\mu\hat k^\perp-
\frac13 k_\perp^2\gamma^{\perp}_\mu\right )\psi^{(1,2,1)}_n (k^2),
\\
\hat \psi^{V(n,2)}_{\mu}(s)=C_{20}^{(n)} \gamma_\mu^\perp\psi^{(1,0,1)}_n (k^2)
                     +C_{22}^{(n)}
\frac{3}{\sqrt{ 2}} \left (k^\perp_\mu\hat k^\perp-
\frac13 k_\perp^2\gamma^{\perp}_\mu\right )\psi^{(1,2,1)}_n (k^2).
 \nn
\eea
But, according to \cite{qq}, we have with a good accuracy
$C_{12}^{(n)}=C_{20}^{(n)}=0$,
so below we put $C_{10}^{(n)}=C_{22}^{(n)}=1$.

We parameterise the $q\bar q$ wave functions of $\rho$, $\omega$
states, $\psi^{(S,L,J)}_{(n)}(k^2)$, with the following formula:
\bea
\label{Ra-1}
\psi^{(S,L,J)}_{(n)}(k^2)=e^{-\beta
|k|^2}\sum\limits_{i=1}^{11} c_i(S,L,J;n) |k|^{i-1}\, ,
\eea
with cutting
parameter  $\beta=1.2 $ GeV$^{-2}$.
In Eq. (\ref{Ra-1}), we use the notation $|k|=\sqrt{s/4-m^2}$ ($m$ is
the mass of the light constituent quark, $m\simeq 350$ MeV).

The constants $c_i(S,L,J;n)$, in GeV units, for
 mesons with $L=0$, $\psi^{(S,L=0,J)}_n(k^2)$, and  
 $L=2$, $\psi^{(S,L=2,J)}_n(k^2)$,
 are presented in Eq. (9) and (10).

\begin{figure}[h]
%Fig. 3
\centerline{\epsfig{file=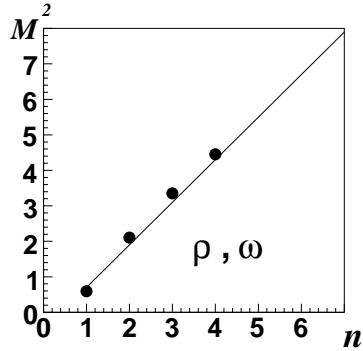,width=5cm}}
\caption{
Trajectory for $\rho_{nS}$ and $\omega_{nS}$ states found in \cite{qq}
 ($M_{\rho(nS)}=M_{\omega(nS)}$). Experimental values of the masses on $\rho$- and
 $\omega$-trajectories are equal to: [$\rho_{1S} (775\pm 10)$, $\rho_{2S}(1460\pm
 20)$, $\rho_{3S}(1870\pm 70)$, $\rho_{4S}(2110\pm 35)$] and
 [$\omega_{1S} (782)$, $\omega_{2S}(1430\pm 50)$,
$\omega_{3S}(\sim1830)$, $\omega_{4S}(2205\pm 40)$].
}
\label{2+1}
\end{figure}

In the solution found in \cite{qq}, the $\rho_{nS}$ and
$\omega_{nS}$ mesons are degenerated: $M_{\rho(nS)}=M_{\omega(nS)}$,
see Fig. \ref{2+1}. Coefficients $c_i(S=1,L=0,J=1;n)$ for $n\le 4 $
(recall that $n$ is radial excitation number) read:
\begin{equation}
\label{Rt-1a} \begin{tabular}{r|r|r|r|r} & $\rho(1S),\omega(1S)$ &
$\rho(2S),\omega(2S)$ & $\rho(3S),\omega(3S)$ & $\rho(4S),\omega(4S)$
\\ \hline $i$ & $\psi_{1}^{(1,0,1)}$ & $\psi_{2}^{(1,0,1)}$ &
$\psi_{3}^{(1,0,1)}$ & $\psi_{4}^{(1,0,1)}$   \\ \hline
 1 &        44.2 &       -47.0 &        34.4 &       256.1  \\
 2 &       147.9 &        96.4 &       367.3 &     -3816.4  \\
 3 &     -2576.7 &      1694.4 &     -6627.1 &     21285.8  \\
 4 &     10145.9 &     -8835.1 &     31300.6 &    -61891.6  \\
 5 &    -20331.5 &     18954.3 &    -72495.7 &    106967.9  \\
 6 &     23805.7 &    -21715.0 &     95497.7 &   -115547.6  \\
 7 &    -16569.8 &     13585.9 &    -73882.6 &     77608.2  \\
 8 &      6338.4 &     -3952.2 &     31633.5 &    -29980.2  \\
 9 &      -941.1 &       119.3 &     -5588.5 &      4927.5  \\
10 &       -59.0 &        26.4 &      -333.1 &       258.1  \\
11 &       -16.0 &        88.7 &        43.2 &       -25.9
\end{tabular}
\end{equation}
For $\rho$ and $\omega$ mesons with dominant $L=2$ we have the
following $c_i(S=1,L=2,J=1;n)$:
\begin{equation} \label{Rt-3}
\begin{tabular}{r|r|r|r|r}
& $\rho(1D),\omega(1D)$ & $\rho(2D),\omega(2D)$ & $\rho(3D),\omega(3D)$
& $\rho(4D),\omega(4D)$  \\ \hline $i$ & $\psi_{1}^{(1,2,1)}$ &
$\psi_{2}^{(1,2,1)}$ & $\psi_{3}^{(1,2,1)}$ & $\psi_{4}^{(1,2,1)}$  \\
\hline
 1 &        32.6 &         1.9 &       295.8 &      1109.3  \\
 2 &      -297.9 &       -20.8 &     -2587.2 &     -9686.9  \\
 3 &      1030.3 &        85.0 &      8635.8 &     32404.0  \\
 4 &     -1720.3 &      -207.3 &    -13721.7 &    -52043.5  \\
 5 &      1257.2 &       242.8 &      9530.7 &     36934.5  \\
 6 &        68.1 &         4.0 &       206.3 &      1219.6  \\
 7 &      -702.1 &      -203.4 &     -4305.9 &    -18749.1  \\
 8 &       419.2 &       125.4 &      2314.3 &     10789.0  \\
 9 &      -113.3 &       -25.0 &      -521.0 &     -2650.0  \\
10 &        68.2 &        16.0 &       378.0 &      1715.0  \\
11 &       -58.4 &       -16.6 &      -340.7 &     -1533.5
\end{tabular}
\end{equation}

\subsubsection{Pion wave function}

For the $\pi(140)$-meson wave function $\psi_{1}^{(0,0,0)}$, the
solution obtained by spectral integral equation is rather satisfactory,
it is given by the coefficients $c_i(S=0,L=0,J=0;n)$ which can be found
in \cite{qq}.

Still,  the pion can be more precisely
described by the wave function found phenomenologically, using the pion
form factor data \cite{g-qq}. The  phenomenological wave function and its parameters
are as follows:
\bea \label{DN-pion}
&&\psi_\pi (s)=c_\pi \bigg(\exp({-b_{1\pi}s})+\beta \exp({-b_{2\pi}s})\bigg),
\nn \\
&& c_\pi= 209.36 {\rm GeV}^{-2},\; b_{1\pi}= 3.57 {\rm GeV}^{-2},\;
b_{2\pi}= 0.4 {\rm GeV}^{-2},\;  \beta =0.01381.
\eea
It should be noted that the difference between the wave function
of Eq. (\ref{DN-pion}) and that found in \cite{qq} is observed either at
rather small relative momenta ($k^2= (s/4-m^2)<0.1$ GeV$^2$) and or at
very large ones.

\subsubsection{Pion emission constant}

The pion--quark coupling $g_\pi$
for the pion emission $q\to \pi +q$ (diagrams of Fig. \ref{2} type) is given by
the quark form factor $g_{\pi (140)\to q\bar q}(s)$ at $s=M^2_\pi$, namely,
$g_\pi=g_{\pi (140)\to q\bar q}(s=M^2_\pi)$. However, the spectral
integral equation does not determine the vertices at $s\leq 4m^2$, so
in our present fit $g_\pi$ is a free parameter.

Describing the widths of $\rho(770)\to \gamma \pi (140)$ and
$\omega(780)\to \gamma \pi (140)$ with the
 use of vector meson (\ref{Rt-3}) and
pion wave functions (\ref{DN-pion}),
we have found two values for the pion bremsstrahlung
coupling:
\bea \label{g-pi}
{\rm Solution\, I}:&& |g_\pi|=16.7 \pm 0.3\ ^{+0.1}_{-2.3}\ , \\
{\rm Solution\, II}:&& |g_\pi|=3.0  \pm 0.3\ ^{+0.1}_{-2.1}\ . \nn
\eea
The pion emission coupling, as is well known, was a
subject of investigation  in  physics of low-energy pion--nucleon
interactions and as well as in nuclear physics. For the pion--nucleon coupling, which
is determined as
$g_{\pi NN}\bigg (\bar \psi\, '_N (\vec \tau \vec \varphi_\pi)i \gamma_5 \psi_N \bigg )$,
the estimations give $g_{\pi NN}^2/(4\pi)\simeq 14$ (see, for example,
\cite{stoks,arndt,bugg-g} and
references therein).

We can turn the description of pion--nucleon vertex into the quark language
using quark model for nucleons:
\be \label{g-pi1}
g_{\pi NN}\bigg (\bar \psi\, '_N (\vec \tau \vec \varphi_\pi)\
i\gamma_5 \psi_N \bigg )
\longrightarrow
g_{\pi qq}\bigg (\bar \psi\, '_q (\vec \tau \vec \varphi_\pi)\
i\gamma_5 \psi_q \bigg )\ ,
\ee
see Appendix A for more detail. In Eq. (\ref{g-pi}), we determine the
vertex $u\to\gamma\pi$
which is a part of the quark-language Lagrangian:
\bea \label{g-pi2}
 g_{\pi qq}\bigg (\bar \psi\, '_q (\vec \tau \vec \varphi_\pi)\
 i\gamma_5 \psi_q \bigg )
&\to & \sqrt{2} g_{\pi qq}\, \varphi^+_{\pi^+}\bigg (
 \bar \psi\, '_d\ i\gamma_5 \psi_u   \bigg )
= g_{\pi }\, \varphi^+_{\pi^+}\bigg (
 \bar \psi\, '_d\ i\gamma_5 \psi_u
  \bigg ), \nn \\
  \sqrt{2} g_{\pi qq}&=& g_{\pi }\, .
\eea
In Appendix A, we show that, making use of the SU(6)-symmetry for nucleons,
one has $ g_{\pi NN}=(5/3)g_{\pi qq}$. So, the SU(6)-symmetry provides us
with $ g_{\pi NN}=(5/3\sqrt{2})g_{\pi}$. It means that Solution I
does not contradict the value $g_{\pi NN}^2/(4\pi)\simeq 14$ 
\cite{stoks,arndt,bugg-g}, thus giving us
$$
16.4\leq g_{\pi NN}^2/(4\pi)\leq 23.2
\; .
$$
Note that in (\ref{g-pi})
we have included systematical errors  which
are due to uncertainties in the reconstruction of wave functions in the fit
\cite{qq}.

\section{Gamma--pion decays of
vector states $V\to\gamma \pi$}

Here  we present  formulae  which are
used below for  $\rho\to \gamma\pi$  and $\omega\to \gamma\pi$ decays.

\subsection{Polarisation  vectors, amplitude and partial width for decays\\
$V\to\gamma \pi$}

Let us introduce notations for the momenta and polarisation  vectors
and  define the amplitudes and decay partial widths.

\subsubsection{Polarisation  vectors of the massive vector particle $V$
and photon}

Polarisations  of the vector meson, $\epsilon^{(V)}_{\mu}$, and of
 virtual photon, $\epsilon^{(\gamma^*)}_\alpha$,  are the transverse
vectors:
\bea
\epsilon^{(V)}_{\beta}p_\beta\ =\ 0\ ,\qquad
\epsilon^{(\gamma^*)}_\alpha q_\alpha\ =\ 0\ ,
\label{pv-1}
\eea
where  $q$ is the virtual photon four-momentum
($q^2\neq 0$) and $p$ is that of the
vector meson ($p^2=M^2_V$).
Polarisation  of the vector meson
obeys the completeness condition as follows:
\bea
 -\sum_{a=1,2,3}\epsilon^{(V)}_\mu(a)
                  \epsilon^{(V) +}_{\mu'}(a)\ =\
g^{\perp V}_{\mu\mu'}\equiv g^{\perp p}_{\mu\mu'}\ ,
\qquad
 g^{\perp p}_{\mu\mu'}\ =\
g_{\mu\mu'}- \frac{p_\mu p_{\mu'}}{p^2}\ ,
\label{pv-2}
\eea
where  $g^{\perp p}_{\mu\mu'}$ is the metric tensor operating in the
space orthogonal to the  momentum $p$.

For virtual photon, $(q^2\neq 0)$, the completeness condition for
polarisation  vectors is written in  three-dimensional space:
\bea
\label{apr6}
-\sum_{a=1,2,3}\epsilon^{(\gamma^*)}_\alpha(a)\,
\epsilon^{(\gamma^*) +}_{\alpha'}(a)=
g^{\perp\gamma^*}_{\alpha\alpha'}\ ,
 \qquad
g^{\perp\gamma^*}_{\alpha\alpha'}\equiv g^{\perp q}_{\alpha\alpha'}=
 g_{\alpha\alpha'}-\frac{q_\alpha q_{\alpha'}}{q^2}\, .
\eea
The polarisation  vector of the real photon $(q^2=0)$ denoted as
$\epsilon^{\gamma}_\alpha$ has two independent components only,
they are orthogonal to the reaction
plane:
\beq \epsilon^{(\gamma)}_\alpha q_\alpha=0\ , \qquad
\epsilon^{(\gamma)}_\alpha p_\alpha=0\ .
\label{apr3}
\eeq
Likewise, the completeness condition for the real photon reads:
\begin{eqnarray} \label{apr4}
&& -\sum_{a=1,2}\epsilon^{(\gamma)}_\alpha(a)
\epsilon^{(\gamma){\bf +}}_{ \alpha'}(a)\ =\
g^{\perp\perp}_{\alpha\alpha'}\ , \\ &&
g^{\perp\perp}_{\alpha\alpha'}\ =\ g_{\alpha\alpha'}- \frac{p_\alpha
p_{\alpha'}}{p^2}-
\frac{q^{\perp}_{\alpha}q^{\perp}_{\alpha'}}{q^2_\perp},\qquad
 q^{\perp}_{\alpha}\equiv g^{\perp V}_{\alpha\alpha'}
 q_{\alpha'}=q_\alpha-\frac{(pq)}{p^2}\,p_\alpha\ .\nonumber
\end{eqnarray}

\subsubsection{Amplitude for the decay $V\to\gamma \pi$}

The decay amplitude $V\to\gamma \pi$
is written as a product of the spin structure and form
factor:
\bea
A_{V\to\gamma \pi}&=& \epsilon_{\alpha}^{(\gamma)}
\epsilon_{\mu}^{(V)}A^{(V\to\gamma \pi)}_{\alpha\mu}\ , \nn \\
A^{(V\to\gamma \pi)}_{\alpha\mu}& =& e\,
S^{(V\to\gamma \pi)}_{\alpha\mu}(p,q) F^{V\to\gamma \pi}(0,M^2_\pi) \ ,
\label{PgV-2}
\eea
with
\bea
S^{(V\to\gamma \pi)}_{\alpha\mu}(p,q)\ =\
\varepsilon_{\alpha\mu  pq}\equiv
\varepsilon_{\alpha\mu\nu_1\nu_2}p_{\nu_1} q_{\nu_2} \ .
\label{PgV-3}
\eea
In (\ref{PgV-2}), the electron charge $e$ is singled out, and in
(\ref{PgV-3}) the tensor $\varepsilon_{\alpha\mu\nu_1\nu_2}$ is the
wholly antisymmetrical. Let us emphasise the specific role of the
spin operator $\varepsilon_{\alpha\mu p q }$. Since
$\varepsilon_{\alpha\mu p p}=0$, this spin operator
is valid for the
reaction with both real ($\gamma$) and virtual ($\gamma^*$) photons, so
Eq. (\ref{PgV-2}) can be used for
the transition with virtual photon,
 with corresponding substitution:
$F^{V\to\gamma \pi}(0)\to F^{V\to\gamma \pi}(q^2)$.

\subsubsection{Partial width for $V\to\gamma \pi$}

The partial width for the decay $V\to\gamma \pi$ is determined as
follows:
\bea
\label{PgV-4}
M_V \Gamma_{V\to\gamma \pi}&=&  \frac13\int d\Phi_2(p;q,p_\pi)\left|
\sum_{\alpha\mu}A^{(V\to\gamma \pi)}_{\alpha\mu}\right|^2 \ =
\nn \\
&=&\frac{\alpha}{24}\frac{(M_V^2-M_\pi^2)^3}{M_V^2}\ |F^{V\to\gamma
\pi}(0,M^2_\pi)|^2\ ,
\nn \\
d\Phi_2(p;q,p_\pi)&=&\frac 12 \frac{d^3q}{(2\pi)^3\, 2q_{0}}
\frac{d^3p_\pi}{(2\pi)^3\, 2p_{\pi 0}} (2\pi)^{4}\delta^{(4)}(p-q
-p_\pi)\ .
\eea
 The summation
is carried out over the photon and vector meson polarisation s, and
$(\varepsilon_{\alpha\mu  pq})^2=(M_V^2-M_\pi^2)^2/2$. In the
final expression $\alpha=e^2/4\pi=1/137$.

\subsection{Double spectral integral representation of the triangle
diagrams with photon emission}

To derive double spectral integral for the form factors with
photon emission by quark and antiquark,
$ F^{V(L)\to\gamma \pi}_{\bigtriangleup^\gamma}(q^2)$ and
$F^{V(L)\to\gamma \pi}_{\bigtriangledown_\gamma}(q^2)$, see Fig.
\ref{1},
 one needs to calculate the double
discontinuities of the triangle diagrams.

\subsubsection{Double
discontinuities of the triangle diagrams}

First,  consider the photon emission by quark, see Fig. \ref{1}a.
Corresponding cuttings for the calculation of  double
discontinuity are shown in Fig. \ref{1}b.

 In the dispersion
representation, the invariant energy in the intermediate state differs
from that in the initial and final states. Because of that, at the
double discontinuity $P\ne p$ and $P'\ne p_\pi$. The following
requirements are imposed on the momenta shown in the diagram of Fig. \ref{1}b
\cite{deut,PR}:
\bea (k_1+k_2)^2\ =\ P^2 \equiv s>4m^2\ ,\qquad (k'_1+k_2)^2\ =\
P'^2\equiv s'>4m^2 \ .
\label{dsi-1}
\eea
The momentum squared of the photon, $q^2$, is fixed:
\bea (p-p_\pi)^2 =(P-P')^2\ =\
(k_1-k_1')^2\ =\ q^2\ .
 \label{dsi-2}
\eea
When  cutting Feynman diagram, the propagators should be
substituted by the residues in the poles. This is equivalent to
the replacement as follows:
$(m^2-k^2_1)^{-1}\to\delta(m^2-k^2_1)$,
$(m^2-k^2_2)^{-1}\to\delta(m^2-k^2_2)$ and
$(m^2-k'^2_1)^{-1}\to\delta(m^2-k'^2_1)$,
so the intermediate-state quarks are mass-on-shell:
\bea
k^2_1=k^2_2 =k'^2_1=m^2.
 \label{dsi-2a}
\eea
Then, for the diagram with  photon
emitted by quark (Fig. \ref{1}a), the double discontinuity
of the amplitude (Fig. \ref{1}b)
 becomes proportional to the three factors:
\bea
\label{dsi-3}
&&{\rm disc}_s{\rm disc}_{s'}A^{V(L)\to\gamma\pi}_{\alpha\mu}(\bigtriangleup^\gamma)\sim
 Z^{V\to\gamma \pi}_{\bigtriangleup^\gamma}
G_{V(L)}(s)G_\pi(s')
\nn\\  &&
\times d\Phi_2(P;k_1,k_2)d\Phi_2(P';k'_1,k'_2)
(2\pi)^32k_{20}\delta^3(\vec k'_2-\vec k_2)
\nn\\ 
&&\times
{\rm Sp}\left[Q^{V(L)}_\mu(k)(\hat k_1+m)Q^{(\gamma)}_\alpha(\hat
k'_1+m)Q^{(\pi)} (-\hat k_2+m) \right] \ .
\eea
The first factor in the right-hand side of (\ref{dsi-3}) consists of the
following vertices: the quark charge factor
$Z^{V\to\gamma\pi}_{\bigtriangleup^\gamma}$
 as well as transition
vertices $V(L)\to q\bar q$ and $\pi\to q\bar q$ which are denoted as
$G_{V(L)}(s)$ and $G_{\pi}(s')$.

The second factor contains space volumes of the two-particle states,
$d\Phi_2(P;k_1,k_2)$ and $d\Phi_2(P';k'_1,k'_2)$,
which correspond to two cuts shown in the diagram of Fig. \ref{1}b (the space volume
is determined in (\ref{PgV-4})). The factor
$(2\pi)^32k_{20}\delta^3(\vec k'_2-\vec k_2)$ takes into account
the fact that one quark line is cut twice.

The third factor in (\ref{dsi-3}) is the trace coming from the summation
over the quark spin states. Since the spin factor in the transition
$V\to q\bar q$ may be of  two types (with dominant $S$- or dominant
$D$-wave), we have the following operators for virtual photon,
$Q^{V(L)}_\mu$, see Eq. (\ref{Rwf-1}):
\bea
\label{dsi-5}
 &&Q^{V(L=0)}_\mu(k) = \gamma^{\perp V}_\mu =\gamma^{\perp P}_\mu
\equiv g^{\perp P}_{\mu\mu'}\gamma_{\mu'} ,
 \nn \\
 &&Q^{V(L=2)}_\mu(k)\ =\ \sqrt{2} \gamma_{\mu'}
X^{(2)}_{\mu'\mu}(k)=
\frac{3}{\sqrt{2}}\left[k_\mu\hat{k}-\frac13 k^2
\gamma^{\perp P}_\mu\right] \,\ ,
\eea
and for the pion:
\be
 Q^{(\pi)}=i\gamma_5\ .
 \label{dsi-5pi}
\ee
Here, $k=(k_1-k_2)/2$ is the relative momentum of the
incoming quarks, $k\perp P=k_1+k_2$, i.e.
$k=k_1^{\perp P}=-k_2^{\perp P}$.

For real photon, we replace:
\bea
\label{dsi-5a}
&&
Q^{(\gamma)}_\alpha \to 
Q^{\perp  \perp }_\alpha \equiv
\gamma^{\perp\perp}_\alpha (P,P')=
g^{\perp\perp}_{\alpha\alpha'}(P,P') \gamma_{\alpha'}\ ,\nn \\
&&
g^{\perp\perp}_{\alpha\alpha'}(P,P')P_{\alpha'}=0, \quad
g^{\perp\perp}_{\alpha\alpha'}(P,P')P'_{\alpha'}=0\ ,
\eea
where $(P-P')^2=0$.
The metric tensor $g^{\perp\perp}_{\alpha\alpha'}(P,P')$  works
in the space orthogonal to the intermediate state momenta:
$g^{\perp\perp}_{\alpha\alpha'}(P,P')= g_{\alpha\alpha'}-
P_{\alpha}P_{\alpha'}/P^2 -
P'^{\perp P}_{\alpha}P'^{\perp P}_{\alpha'}/P'^{\perp
P}_{\alpha''}P'^{\perp P}_{\alpha''} $.

Actually, for the real photon  we can use  simpler oprator, say,
$Q^{(\gamma)}_\alpha= \gamma_\alpha^\perp$, because in the considered decay we
should have the same result for both choices, $Q^{(\gamma)}_\alpha$ or
$Q^{\perp  \perp }_\alpha$,
due to the  spin operator structure (\ref{PgV-3}).
However, here we use (\ref{dsi-5a}) to emphasise an important point
for this type of reactions: the amplitude for transversely polarized photons
is determined by the spectral integral with  transversely polarized photons
in the intermediate states as well.

For the photon emission, there are two diagrams: the second one is
similar to that of  Fig. \ref{1}a but with the
emission of  photon by antiquark, it is shown in Fig. \ref{1}c.
The double discontinuity of the
corresponding  amplitude is determined by cuttings shown in
Fig. \ref{1}d:
\bea
\label{dsi-5b}
&&{\rm disc}_s{\rm disc}_{s'}A^{V(L)\to\gamma
\pi}_{\alpha\mu}(\bigtriangledown_\gamma)\sim
Z_{V\to\gamma \pi}(\bigtriangledown_\gamma)
G_{V(L)}(s)G_\pi(s')
 \nn \\
&&\times
d\Phi_2(P;k_1,k_2)d\Phi_2(P';k'_1,k'_2)
(2\pi)^32k_{10}\delta^3(\vec k'_1-\vec k_1)
\nn \\ 
&&\times
{\rm Sp}\left[Q^{V(L)}_\mu(k)(\hat k_1+m)Q^{(\pi)}(-\hat
k'_2+m)Q^{(\gamma)}_\alpha (-\hat k_2+m) \right] \ .
\eea
Likewise, there are two traces for  two
transitions with photon emission by  quark and antiquark:
\bea
\label{dsi-6}
&& Sp^{V(L)\to\gamma
\pi}_{\alpha\mu}(\bigtriangleup^\gamma)
=
-{\rm Sp}\left[Q^{V(L)}_\mu(k)(\hat k_1+m)Q^{(\gamma)}_\alpha(\hat
k'_1+m)Q^{(\pi)} (-\hat k_2+m) \right] \ ,
 \nn \\
&&  Sp^{V(L)\to\gamma
\pi}_{\alpha\mu}(\bigtriangledown_\gamma)
=
-{\rm Sp}\left[Q^{V(L)}_\mu(k)(\hat k_1+m)Q^{(\pi)}(-\hat
k'_2+m)Q^{(\gamma)}_\alpha (-\hat k_2+m) \right] .\nn\\
\eea
To calculate the invariant form factors
$F^{V(L)\to\gamma \pi}_{\bigtriangleup^\gamma}(0)$
and
$F^{V(L)\to\gamma \pi}_{\bigtriangledown_\gamma}(0)$,
we
should extract from (\ref{dsi-6}) the intermediate-state spin operator:
\bea
S^{(V\to\gamma \pi)}_{\alpha\mu}(P,\widetilde q)\ =\
 \varepsilon_{\alpha\mu P\widetilde q }\ , \qquad \widetilde q=P-P'\ .
\label{dsi-7}
\eea
Therefore, we have:
\bea
\label{dsi-81}
 Sp^{V(L)\to\gamma
\pi}_{\alpha\mu}(\bigtriangleup^\gamma)
&=& S^{(V\to\gamma \pi)}_{\alpha\mu}(P,\widetilde q)
S^{V(L)\to\gamma
\pi}_{\bigtriangleup^\gamma}(s,s',q^2)\ ,
\nn \\
 Sp^{V(L)\to\gamma
\pi}_{\alpha\mu}(\bigtriangledown_\gamma)
&=& S^{(V\to\gamma \pi)}_{\alpha\mu}(P,\widetilde q)
S^{V(L)\to\gamma
\pi}_{\bigtriangledown_\gamma}(s,s',q^2)\ ,
\eea
where
\bea
\label{dsi-91}
\frac{\left( Sp^{V(L)\to\gamma
\pi}_{\alpha\mu}(\bigtriangleup^\gamma)
 S^{(V\to\gamma \pi)}_{\alpha\mu}(P,\widetilde q)\right)}
{\left(S^{(V\to\gamma \pi)}_{\alpha\mu}(P,\widetilde q)\right)^2} &=&
S^{V(L)\to\gamma
\pi}_{\bigtriangleup^\gamma}(s,s',q^2)\ ,
\nn \\
\frac{\left( Sp^{V(L)\to\gamma
\pi}_{\alpha\mu}(\bigtriangledown_\gamma)
 S^{(V\to\gamma \pi)}_{\alpha\mu}(P,\widetilde q)\right)}
{\left(S^{(V\to\gamma \pi)}_{\alpha\mu}(P,\widetilde q)\right)^2} &=&
S^{V(L)\to\gamma
\pi}_{\bigtriangledown_\gamma}(s,s',q^2)\ .
\eea
Taking into account the expression
${\rm Sp}[\gamma_5\gamma_{\alpha_1}\gamma_{\alpha_2}\gamma_{\alpha_3}\gamma_{\alpha_4}]=
4i\varepsilon_{\alpha_1\alpha_2\alpha_3\alpha_4}$
we obtain:
\bea
\label{dsi-10}
 S^{V(L=0)\to\gamma
\pi}_{\bigtriangleup^\gamma}(s,s',q^2)=
S^{V(0)\to\gamma
\pi}_{\bigtriangledown_\gamma}(s,s',q^2)= -4m\ ,
\\ \nn 
S^{V(L=2)\to\gamma
\pi}_{\bigtriangleup^\gamma}(s,s',q^2)=
S^{V(2)\to\gamma
\pi}_{\bigtriangledown_\gamma}(s,s',q^2)=
-\frac{m}{\sqrt{2}}
\left[2m^2+\!s+\!\frac{6ss'q^2}{\lambda(s,s',q^2)}\right] ,
\eea
with
\bea
\lambda (s,s',q^2)=(s-s')^2-2q^2(s+s')+q^4 .
\label{dsi-11}
\eea
The photon emission amplitude, being determined by two diagrams of Fig.
\ref{1}a and Fig. \ref{1}c, reads
\bea
A^{V(L)\to\gamma \pi}_{(\bigtriangleup^\gamma +\bigtriangledown_\gamma)\alpha\mu}
\!=\! e\,
\varepsilon_{\alpha\mu p q}\!\left[
Z^{V\to\gamma \pi}_{\bigtriangleup^\gamma}
 \!F^{V(L)\to\gamma \pi}_{\bigtriangleup^\gamma}(q^2,M^2_\pi)\! + \!
Z^{V\to\gamma \pi}_{\bigtriangledown_\gamma}
 F^{V(L)\to\gamma \pi}_{\bigtriangledown_\gamma}(q^2,M^2_\pi) \right]\! ,\nn \\
\label{dsi-12}
\eea
while the double discontinuities of the form factors in (\ref{dsi-12})
are equal to:
\bea  \label{dsi-13}
&&
{\rm disc}_s {\rm disc}_{s'} F^{V(L)\to\gamma \pi}_{\bigtriangleup^\gamma}
(q^2,M^2_\pi)
= G_{V(L)}(s)G_\pi(s')\nn \\
&&\times
d\Phi_2(P;k_1,k_2)d\Phi_2(P';k'_1,k'_2) (2\pi)^32k_{20}\delta^3(\vec
k'_2-\vec k_2) S^{V(L)\to\gamma
\pi}_{\bigtriangleup^\gamma}(s,s',q^2) ,
\nn \\
&&{\rm disc}_s {\rm disc}_{s'}
F^{V(L)\to\gamma \pi}_{\bigtriangledown_\gamma}(q^2,M^2_\pi)=
G_{V(L)}(s)G_\pi(s')
\nn  \\
&&\times  d\Phi_2(P;k_1,k_2)d\Phi_2(P';k'_1,k'_2)
(2\pi)^32k_{10}\delta^3(\vec k'_1-\vec k_1)
S^{V(L)\to\gamma
\pi}_{\bigtriangledown_\gamma}(s,s',q^2) .
\eea

\subsubsection{The double spectral integral for the form factors with
photon emission by quark and antiquark}

The equation (\ref{dsi-13}) defines the form factor
through the dispersion integral as follows:
\bea
&& F^{V(L)\to\gamma \pi}_{\bigtriangleup^\gamma}(q^2,M^2_\pi)
= \int\limits^{\infty}_{4m^2}
\frac{ds}{\pi} \int\limits^{\infty}_{4m^2}\frac{ds'}{\pi}
\frac{{\rm disc}_s {\rm disc}_{s'}  F^{V(L)\to\gamma \pi}_{\bigtriangleup^\gamma}
(q^2,M^2_\pi)}
{(s-M^2_{V(L)})(s'-M^2_\pi)} \ , \nn \\
&&F^{V(L)\to\gamma \pi}_{\bigtriangledown_\gamma}(q^2,M^2_\pi)
= \int\limits^{\infty}_{4m^2}
\frac{ds}{\pi} \int\limits^{\infty}_{4m^2}\frac{ds'}{\pi}
\frac{{\rm disc}_s {\rm disc}_{s'}  F^{V(L)\to\gamma \pi}_{\bigtriangledown^\gamma}
(q^2,M^2_\pi)}
{(s-M^2_{V(L)})(s'-M^2_\pi)} \ .
\label{dsi-14}
\eea
We have
\be  \label{dsi-14a}
 F^{V(L)\to\gamma \pi}_{\bigtriangleup^\gamma}(q^2,M^2_\pi)=
F^{V(L)\to\gamma \pi}_{\bigtriangledown_\gamma}(q^2,M^2_\pi)
\ee
at equal masses of the quark and antiquark  -- 
 just this case is  considered here.
In (\ref{dsi-14}), we omit
subtraction terms, assuming that  the convergence of
(\ref{dsi-14}) is guaranteed by the vertices $G_{V(L)}(s)$
and $G_\pi(s')$. Furthermore, we define the wave
functions of the $q\bar q$ systems:
$ \psi_{V(L)}(s) = G_{V(L)}(s)/(s-M_{V(L)}^2)$
and $\psi_\pi(s') = G_\pi(s')/(s'-M_\pi^2)$.

After
integrating over the momenta in accordance with (\ref{dsi-13}), one can
represent (\ref{dsi-14}) in the following form:
 \bea \label{dsi-16}
 F^{V(L)\to\gamma \pi}_{\bigtriangleup^\gamma}(q^2,M^2_\pi) &=&
F^{V(L)\to\gamma \pi}_{\bigtriangledown_\gamma}(q^2,M^2_\pi)=
 \int \limits_{4m^2}^\infty
\frac{dsds'}{16\pi^2} \psi_{V(L)}(s)\psi_{\pi}(s')
 \nn\\
&\times&\frac{\Theta\left(-ss'q^2-m^2\lambda(s,s',q^2)\right)}
{\sqrt{\lambda(s,s',q^2)}}
S^{V(L)\to\gamma \pi}_{\bigtriangleup^\gamma}(s,s',q^2)\ ,
\eea
where $\Theta(X)$ is the  step-function: $\Theta(X)=1$ at
$X\ge 0$ and $\Theta(X)=0$ at $X<0$.

\subsubsection{Z-factors for photon emission }

\begin{figure}[h]
%\Fig. 3
\centerline{\epsfig{file=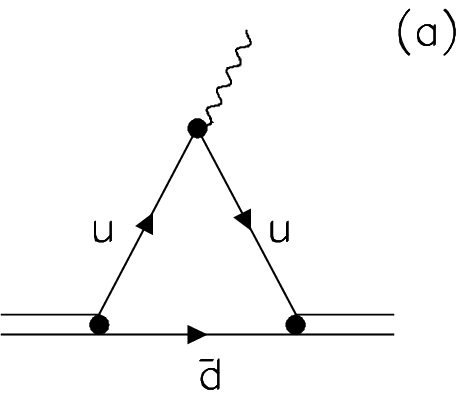,width=3cm}\hspace{1cm}
            \epsfig{file=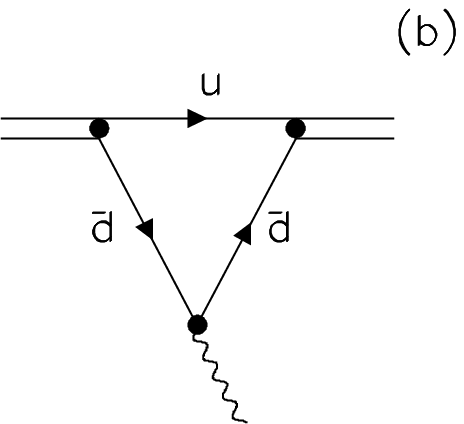,width=3cm}}
\caption{Diagrams for the determination of
Z-factors in the reaction $ \rho^+ \to \gamma \pi^+ $ with photon
 emission .}
\label{3}
\end{figure}

For the $\rho^+$ meson, the photon emission is determined by two
diagrams, see Figs. \ref{3}a and \ref{3}b, which give us
the following charge factors:
\be \label{dsi-17}
Z^{\rho^+\to\gamma \pi^+}_{\bigtriangleup^\gamma}=e_u=\frac 23, \qquad
Z^{\rho^+\to\gamma \pi^+}_{\bigtriangledown_\gamma}=e_d=-\frac 13\ .
\ee

\begin{figure}[h]
%\Fig. 4
\centerline{\epsfig{file=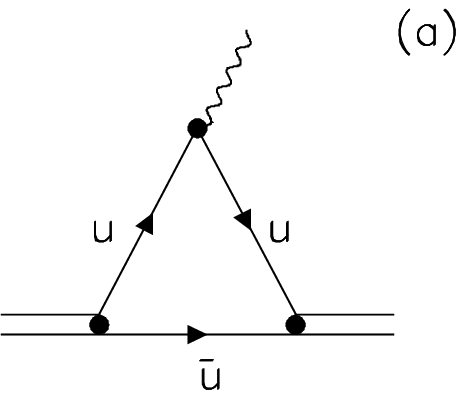,width=3cm}\hspace{1cm}
            \epsfig{file=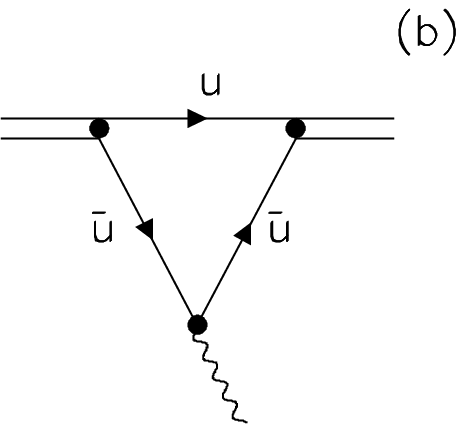,width=3cm}\hspace{1cm}
            \epsfig{file=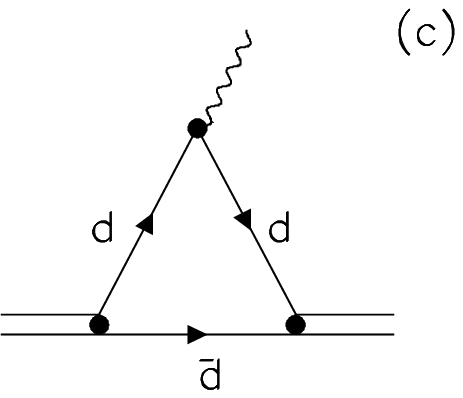,width=3cm}\hspace{1cm}
            \epsfig{file=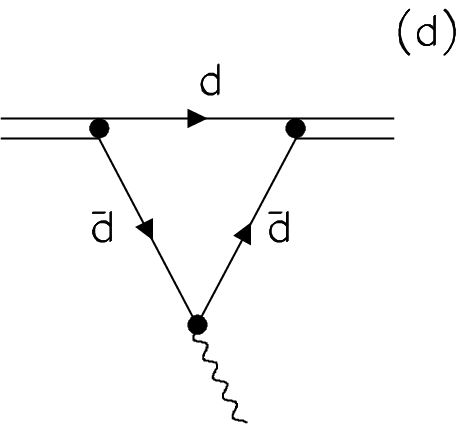,width=3cm}}
\caption{Diagrams for the determination of  Z-factors in
 the reactions $\rho^0\to\gamma\pi^0$ and
     $\omega^0\to\gamma\pi^0$ with photon emission.}
\label{4}
\end{figure}

For neutral vector mesons ($\rho^0$, $\omega$), we have four diagrams,
see Fig. \ref{4}, which
result in the charge factors as follows:
\bea  \label{dsi-18}
&& Z^{\rho^0\to\gamma \pi^0}_{\bigtriangleup^\gamma}=
Z^{\rho^0\to\gamma \pi^0}_{\bigtriangledown_\gamma}=
\frac 12 (e_u+e_d)=\frac 16\ , \nn \\
&& Z^{\omega\to\gamma \pi^0}_{\bigtriangleup^\gamma}=
Z^{\omega\to\gamma \pi^0}_{\bigtriangledown_\gamma}=
\frac 12 (e_u-e_d)=\frac 12\ .
\eea
In (\ref{dsi-18}), we use the standard flavour wave functions for
$(I=1,I_3=0)$ and $(I=0,I_3=0)$ states:
$\rho^0=\pi^0=(u\bar u-d\bar d)/\sqrt 2$  and
$\omega=(u\bar u+d\bar d)/\sqrt 2$.

\subsubsection{Decay form factors at  $Q^2=-q^2\to 0$}

To calculate the integral at small $Q^2$,
we  substitute:
\be
\label{dsi-16a}
s=\Sigma+\frac 12 zQ,\quad
s'=\Sigma-\frac 12 zQ,\quad
q^2=-Q^2\, .
\ee
In the region $Q^2\ll 4m^2$, the form factors (\ref{dsi-16}) can be
written as
\bea
 F^{V(L)\to\gamma \pi}_{\bigtriangleup^\gamma}(-Q^2,M^2_\pi) &=&
F^{V(L)\to\gamma \pi}_{\bigtriangledown_\gamma}(-Q^2,M^2_\pi)=
\int \limits_{4m^2}^\infty
\frac{d\Sigma }{\pi} \psi_{V(L)}(\Sigma)\psi_\pi(\Sigma)
\nn \\
&\times&\int \limits_{-b}^{+b}\frac{ dz}{\pi}\;
\frac {S^{V(L)\to\gamma \pi}_{\bigtriangleup^\gamma}
(\Sigma+\frac 12 zQ,\Sigma-\frac 12 zQ,-Q^2)}
{16\sqrt{\Lambda(\Sigma ,z ,Q^2)}}\ ,\nn
\eea
\be
\label{dsi-16b}
b=\sqrt{\Sigma (\frac{\Sigma}{m^2} -4)}, \qquad
\Lambda(\Sigma ,z, Q^2)=(z^2+4\Sigma) Q^2 \ .
\ee
After integrating over $z$ and substituting $\Sigma\to s$, the
form factors for $L=0,2$ read:
\bea \nn
F^{V(0)\to\gamma \pi}_{\bigtriangleup^\gamma}(0,M^2_\pi)& =& 
F^{V(0)\to\gamma \pi}_{\bigtriangledown_\gamma}(0,M^2_\pi)=
-4m
\int \limits_{4m^2}^\infty \frac{ds}{16\pi^2}
\psi_{\pi}(s)\psi_{V(0)}(s)\nn\\
&\times&\ln{\frac{s+\srho}{s-\srho}} \ ,
 \nn \\
 F^{V(2)\to\gamma \pi}_{\bigtriangleup^\gamma}(0,M^2_\pi) &=&
F^{V(2)\to\gamma \pi}_{\bigtriangledown_\gamma}(0,M^2_\pi)=
-m/\sqrt{2}
\int \limits_{4m^2}^\infty \frac{ds}{4\pi^2}
\psi_{\pi}(s)\psi_{V(2)}(s)
\nn\\
&\times&\left[(2m^2+s)
\ln\frac{\sqrt{s}+\sqrt{s-4m^2}}{\sqrt{s}-\sqrt{s-4m^2}}
+3\sqrt{s(s-4m^2)}\right] \! .
\label{PgV31}
\eea
 Remind that wave functions $\psi_{V(0)}(s)=\psi^{(1,0,1)}_n (k^2)$,
$\psi_{V(2)}(s)=\psi^{(1,2,1)}_n (k^2)$ and
$\psi_{\pi}(s)$  are presented in Section I.

\subsubsection{Normalisation  conditions for the wave functions
$\psi_{\pi}(s)$ and $\psi_{V(L=0,2)}(s)$}

It is convenient to write the normalisation  conditions
for  $\psi_{\pi}(s)$ and $\psi_{V(L)}(s)$
using the charge form factor
of a meson:
\bea
F_{charge}(0)\ =\ 1\ .
 \label{norm-1}
\eea
The amplitude of the charge factor is defined by the photon-emission
triangle diagram
with
$(q\bar q)_{in}=(q\bar q)_{out}$.
For the pion,
the amplitude takes the form:
\bea
A_\alpha(q)\ =\
e(p+p_\pi)_\alpha F_{charge}(q^2)\ ,
\label{norm-2}
\eea
while $F_{charge}(q^2)$ can be calculated in the same way as the
transition form factors considered above.
The normalisation  condition for
pion reads:
\bea
\label{norm-4}
1&=&\int\limits_{4m^2}^{\infty}\frac{ds}{16\pi^2}\ \psi_\pi^2(s)\ 2s\
\sqrt{\frac{s-4m^2}{s}}\ .
\eea
For  vector meson  $V(L)$, the normalisation  condition may be determined
by
averaging over spins of the massive vector
particle, see \cite{book3,BS,EPJA,YF} for detail. Then,
 the normalisation  condition reads:
 \bea
\label{norm-5}
1&=&\frac13\int\limits_{4m^2}^{\infty}
\frac{ds}{16\pi^2}\ \psi_{V(0)}^2(s)\ 4\left(s+2m^2\right)
\sqrt{\frac{s-4m^2}{s}}\ ,
\nn \\
1&=&\frac13\int\limits_{4m^2}^{\infty}
\frac{ds}{16\pi^2}\ \psi_{V(2)}^2(s)\ \frac{(8m^2+s)(s-4m^2)^2}{8}
\sqrt{\frac{s-4m^2}{s}}\ .
\eea
Recall that here
 $\psi_{V(0)}(s)=\psi^{(1,0,1)}_n (k^2)$ and
$\psi_{V(2)}(s)=\psi^{(1,2,1)}_n (k^2)$ with $k^2=s/4\, -m^2$ .

\subsubsection{Vector mesons: normalisation  condition in case of two-component
wave functions}
In the solution found in \cite{qq}, the wave functions $\psi_{V(0)}$ and
$\psi_{V(2)}$ are orthogonal to each other with a good accuracy.
 Generally, vector states  may mix.
 Then the vector mesons have two-component wave functions,
 see (\ref{Rwf-1mix}), and
normalisation  condition reads:
 \bea
\label{norm-vector2}
\delta_{ab}&=&\frac13\int\limits_{4m^2}^{\infty}
\frac{ds}{16\pi^2}\ C_{a0}^{(n)}C_{b0}^{(n)}\bigg( \psi^{(1,0,1)}_n (k^2)\bigg )^2
 4\left(s+2m^2\right)
\sqrt{\frac{s-4m^2}{s}}\
\nn \\
 &+&\frac13\int\limits_{4m^2}^{\infty}
\frac{ds}{16\pi^2}\ \bigg(C_{a0}^{(n)}C_{b2}^{(n)}+C_{b0}^{(n)}C_{a2}^{(n)}\bigg )
\psi^{(1,0,1)}_n (k^2)\psi^{(1,2,1)}_n (k^2)
\nn \\ &&\times\sqrt{2}\ \frac{(s-4m^2)^2}{6}
\sqrt{\frac{s-4m^2}{s}}\
\nn \\
 &+&\frac13\int\limits_{4m^2}^{\infty}
\frac{ds}{16\pi^2}\  C_{a2}^{(n)}C_{b2}^{(n)}\bigg( \psi^{(1,2,1)}_n (k^2)\bigg )^2\
\nn \\&&\times\frac{(8m^2+s)(s-4m^2)^2}{8}
\sqrt{\frac{s-4m^2}{s}}\ .
\eea

\section{Double spectral integral representation of the triangle
diagrams with pion emission}

Here, we calculate the double spectral integral for the transition form
factors with the emission of pion by quark,
$ F^{V(L)\to\gamma \pi}_{\bigtriangleup^\pi}(0,M^2_\pi)$
(diagram of Fig. \ref{2}a)
and antiquark,
$F^{V(L)\to\gamma \pi}_{\bigtriangledown_\pi}(0,M^2_\pi)$
(diagram of Fig. \ref{2}c).

\subsubsection{Double
discontinuities of the triangle diagrams}

For the diagram of Fig. \ref{2}a, the cuttings
are shown in Fig. \ref{2}b, with the following notations:
\bea \label{dsi-1p}
&& k^2_1=k^2_2 =k'^2_1=m^2, \nn \\
&& (k_1+k_2)^2\ =\ P^2 \equiv s>4m^2\ ,\qquad (k'_1+k_2)^2\ =\
P'^2\equiv s'>4m^2, \nn \\
&& (P-P')^2\ =\
(k_1-k'_1)^2 \ = \ p_\pi^2 =M^2_\pi\ .
\eea
For the diagram of Fig. \ref{2}a, the double discontinuity, determined by
Fig. \ref{2}b, contains three factors:
\bea
\label{dsi-3p}
&&{\rm disc}_s {\rm disc}_{s'}A^{V(L)\to\gamma
\pi}_{\alpha\mu}(\bigtriangleup^\pi)\sim 
Z_{V\to\gamma \pi}(\bigtriangleup^\pi)
g_\pi G_{V(L)}(s)G_\gamma(s') \nn \\ \nn
& &\times
d\Phi_2(P;k_1,k_2)d\Phi_2(P';k'_1,k'_2)
(2\pi)^32k_{20}\delta^3(\vec k'_2-\vec k_2)
\\
&&\times 
{\rm Sp}\left[Q^{V(L)}_\mu(k)(\hat k_1+m)Q^{(\pi)}(\hat
k'_1+m)Q^{(\gamma_\perp)}_\alpha (-\hat k_2+m) \right] \ .
\eea
The right-hand side of (\ref{dsi-3p}) is determined by the
the quark charge factor $Z_{V\to\gamma
\pi}(\bigtriangleup^\pi)$,
the transition
vertices $V(L)\to q\bar q$ and $\gamma\to q\bar q$
 and pion--quark coupling $g_\pi$.
The trace in (\ref{dsi-3p})
contains the operators
$Q^{(\pi})$  and $Q^{(\gamma_\perp)}_\alpha$ which are determined in
(\ref{dsi-5a}):
$Q^{(\gamma_\perp)}_\alpha=
\gamma^{\perp\perp}_\alpha (P,P')$ and $ Q^{(\pi)}=i\gamma_5$.

The diagram with the emission of pion by antiquark is shown in Fig. \ref{2}c.
The double discontinuity
of the corresponding  amplitude, Fig. \ref{2}d, is written similarly to
(\ref{dsi-3p}). We have:
\bea
\label{dsi-5p}
 &&{\rm disc}_s{\rm disc}_{s'}A^{V(L)\to\gamma
\pi}_{\alpha\mu}(\bigtriangledown_\pi)\sim
Z_{V\to\gamma \pi}(\bigtriangledown_\pi)
g_\pi G_{V(L)}(s)G_\gamma(s')\nn \\ \nn
&&\times 
d\Phi_2(P;k_1,k_2)d\Phi_2(P';k'_1,k'_2)
(2\pi)^32k_{10}\delta^3(\vec k'_1-\vec k_1)
\\
&&\times 
{\rm Sp}\left[Q^{V(L)}_\mu(k)(\hat k_1+m)Q^{(\gamma_\perp)}_\alpha(-\hat
k'_2+m)Q^{(\pi)} (-\hat k_2+m) \right] \ .
\eea
Correspondingly, we have two traces for  two
transitions with pion emission by the quark and antiquark:
\bea
\label{dsi-6p}
 Sp^{V(L)\to\gamma \pi}_{\alpha\mu}(\bigtriangleup^\pi)
&=&
\!-\!{\rm Sp}\left[Q^{V(L)}_\mu(k)(\hat k_1+m)Q^{(\pi)}(\hat
k'_1+m)Q^{(\gamma_\perp)}_\alpha (-\hat k_2+m) \right]
 \nn \\
&=& S^{(V\to\gamma \pi)}_{\alpha\mu}(P,P-P')
S^{V(L)\to\gamma
\pi}_{\bigtriangleup^\pi}(s,s',(P-P')^2) \ ,
 \nn \\
 Sp^{V(L)\to\gamma \pi}_{\alpha\mu}(\bigtriangledown_\pi)
&=&
\!-\!{\rm Sp}\left[Q^{V(L)}_\mu(k)(\hat k_1\!+\!m)Q^{(\gamma_\perp)}_\alpha
(-\hat k'_2+m)Q^{(\pi)} (-\hat k_2+m) \right]  \nn \\
&=& S^{(V\to\gamma \pi)}_{\alpha\mu}(P,P-P')
S^{V(L)\to\gamma
\pi}_{\bigtriangledown_\pi}(s,s',(P-P')^2)\, ,\nn\\
&&S^{(V\to\gamma \pi)}_{\alpha\mu}(P,P-P') =
  \varepsilon_{\alpha\mu P(P-P') }\ =\  - \varepsilon_{\alpha\mu PP' }\ .
\eea
Here,
\bea
\label{dsi-91p}
\frac{\left(Sp^{V(L)\to\gamma
\pi}_{\alpha\mu}(\bigtriangleup^\pi)
 S^{(V\to\gamma \pi)}_{\alpha\mu}(P,P-P')\right)}
{\left(S^{(V\to\gamma \pi)}_{\alpha\mu}(P,P-P')\right)^2} &=&
S^{V(L)\to\gamma \pi}_{\bigtriangleup^\pi}(s,s',(P-P')^2) \ ,
\nn \\
\frac{\left( Sp^{V(L)\to\gamma \pi}_{\alpha\mu}(\bigtriangledown_\pi)
 S^{(V\to\gamma \pi)}_{\alpha\mu}(P,P-P')\right)}
{\left(S^{(V\to\gamma \pi)}_{\alpha\mu}(P,P-P')\right)^2} &=&
S^{V(L)\to\gamma \pi}_{\bigtriangledown_\pi}(s,s',(P-P')^2) \ .
\eea
As a result, we obtain:
\bea
\label{dsi-10p}
S^{V(0)\to\gamma \pi}_{\bigtriangleup^\pi}(s,s',(P-P')^2)
&=&
S^{V(0)\to\gamma \pi}_{\bigtriangledown_\pi}(s,s',(P-P')^2)= 4m\ ,
 \nn \\
S^{V(2)\to\gamma \pi}_{\bigtriangleup^\pi}(s,s',(P-P')^2)
&=&
S^{V(2)\to\gamma
\pi}_{\bigtriangledown_\pi}(s,s',(P-P')^2)
\nn \\
&=&\frac{m}{\sqrt{2}}
\left[2m^2+s+\frac{6ss'(P-P')^2}{\lambda(s,s',(P-P')^2)}\right]\ .
\eea
Let us note that spin factors
$ S^{V(0)\to\gamma \pi}_{\bigtriangleup^\pi}(s,s',(P-P')^2)$
and $S^{V(2)\to\gamma \pi}_{\bigtriangleup^\pi}(s,s',(P-P')^2)$
differ by the sign only from those for photon emission
$ S^{V(0)\to\gamma \pi}_{\bigtriangleup^\gamma}(s,s',q^2)$
and $S^{V(2)\to\gamma \pi}_{\bigtriangleup^\gamma}(s,s',q^2)$, given
by (\ref{dsi-10}).
The pion emission amplitude,
 considered as a function
of $q^2$ and $p_\pi^2$, is determined
by two processes (Figs. \ref{2}a, \ref{2}c):
\bea  \label{dsi-11p}
A^{(V(L)\to\gamma
\pi)}_{\alpha\mu} (\bigtriangleup^\pi +\bigtriangledown_\pi)
 = e\,
\varepsilon_{\alpha\mu p q}\nn\\
\times\left[
Z^{V\to\gamma \pi}_{\bigtriangleup^\pi}
 F^{V(L)\to\gamma \pi}_{\bigtriangleup^\pi}(q^2,p_\pi^2) +
Z^{V\to\gamma \pi}_{\bigtriangledown_\pi}
 F^{V(L)\to\gamma \pi}_{\bigtriangledown_\pi}(q^2,p_\pi^2) \right] ,
\eea
with
\be \label{dsi-12p}
 F^{V(L)\to\gamma \pi}_{\bigtriangleup^\pi}(q^2,p_\pi^2) =
F^{V(L)\to\gamma \pi}_{\bigtriangledown_\pi}(q^2,p_\pi^2)
\ee
due to the equality (\ref{dsi-10p})
\be
{\rm disc}_s {\rm disc}_{s'} F^{V(L)\to\gamma \pi}_{\bigtriangleup^\pi}(s',p_\pi^2)=
{\rm disc}_s {\rm disc}_{s'}
F^{V(L)\to\gamma \pi}_{\bigtriangledown_\pi}(s',p_\pi^2).
\ee

\subsubsection{
The double spectral integral for the form factors with
pion emission}

The form factors read:
\bea \label{dsi-14p}
F^{V(L)\to\gamma \pi}_{\bigtriangleup^\pi}(q^2,p_\pi^2)& =&
F^{V(L)\to\gamma \pi}_{\bigtriangledown_\pi}(q^2,p_\pi^2)
  \nn \\
 & =&\int\limits^{\infty}_{4m^2}
\frac{ds}{\pi} \int\limits^{\infty}_{4m^2}\frac{ds'}{\pi}
\frac{{\rm disc}_s {\rm disc}_{s'}  F^{V(L)\to\gamma \pi}_{\bigtriangleup^\pi}
(s',p_\pi^2)}
{(s-M^2_{V(L)})(s'-q^2)} \ .
\eea
As in (\ref{dsi-14}), we
assume that  the convergence of
(\ref{dsi-14p}) is guaranteed by the vertices $G_{V(L)}(s)$
and $G_\gamma(s')$.

Futhermore, we
consider the production of
photon, $q^2=0$, and use the photon wave function
$\psi_{\gamma}(s') =G_\gamma(s')/s'$.
After integrating over intermediate-state quark  momenta, one
can represent (\ref{dsi-14p}) for $p_\pi^2\leq 0$ in the following form:
 \bea \label{dsi-16p}
F^{V(L)\to\gamma \pi}_{\bigtriangleup^\pi}(0,p_\pi^2) &=&
F^{V(L)\to\gamma \pi}_{\bigtriangledown_\pi}(0,p_\pi^2)\nn\\ \nn
&=& g_\pi \int \limits_{4m^2}^\infty
\frac{dsds'}{16\pi^2} \psi_{V(L)}(s)\psi_{\gamma}(s')
\nn\\
&\times&\frac{\Theta\left(-ss'p_\pi^2-m^2\lambda(s,s',p_\pi^2)\right)}
{\sqrt{\lambda(s,s',p_\pi^2)}}
S^{V(L)\to\gamma \pi}_{\bigtriangleup^\pi}(s,s',p_\pi^2) .
\eea
The  step-function $\Theta(X)$ was defined in (\ref{dsi-16}).

Let us emphasise once again
that Eq. (\ref{dsi-16p}) is valid in the region
$p_\pi^2\leq 0$ only. To obtain form factors at $p_\pi^2 =M^2_\pi$, one needs
to continue Eq. (\ref{dsi-16p}) to the region $p_\pi^2 > 0$. Since the
form factors are analytical functions in the vicinity of $p_\pi^2 = 0$, the
straightforward way is to expand them in a series over $p_\pi^2 $ keeping
constant and linear terms only:
\be   \label{dsi-16pa}
 F^{V(L)\to\gamma \pi}_{\bigtriangleup^\pi}(0,p_\pi^2) =
 F^{V(L)\to\gamma \pi}_{\bigtriangleup^\pi}(0,0)+
 p_\pi^2 \frac{d}{dp_\pi^2} F^{V(L)\to\gamma \pi}_{\bigtriangleup^\pi}(0,0) \ .
\ee
One can approximate
$p_\pi^2\cdot  F^{V(L)\to\gamma \pi}_{\bigtriangleup^\pi}(0,0)/dp_\pi^2=
 F^{V(L)\to\gamma \pi}_{\bigtriangleup^\pi}(0,0)-
F^{V(L)\to\gamma \pi}_{\bigtriangleup^\pi}(0,-M^2_\pi)$
(here $p_\pi^2=-M^2_\pi$). Then
\bea   \label{dsi-16pb}
 F^{V(L)\to\gamma \pi}_{\bigtriangleup^\pi}(0,M^2_\pi)
 &=& F^{V(L)\to\gamma \pi}_{\bigtriangledown_\pi}(0,M^2_\pi)\nn \\
 & =&
2 F^{V(L)\to\gamma \pi}_{\bigtriangleup^\pi}(0,0)-
F^{V(L)\to\gamma \pi}_{\bigtriangleup^\pi}(0,-M^2_\pi)\ .
\eea
Both form factors,
$ F^{V(L)\to\gamma \pi}_{\bigtriangleup^\pi}(0,0)$ and
$F^{V(L)\to\gamma \pi}_{\bigtriangleup^\pi}(0,-M^2_\pi)$, are
calculated according to Eq. (\ref{dsi-16p}).

\subsubsection{ Z-factors for pion emission}

\begin{figure}[h]
%Fig. 6
\vspace{-0.5cm}
\centerline{\epsfig{file=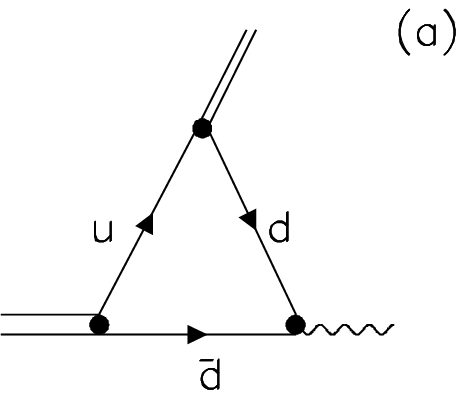,width=3cm}\hspace{1cm}
            \epsfig{file=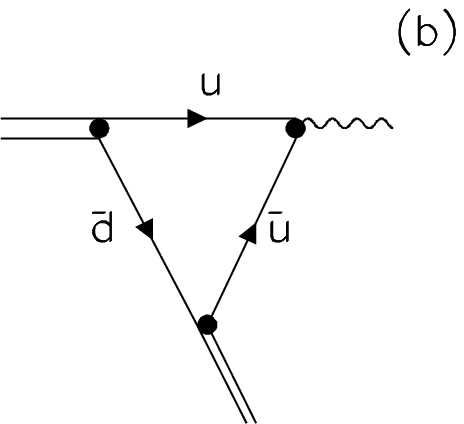,width=3cm}\hspace{1cm}
            \epsfig{file=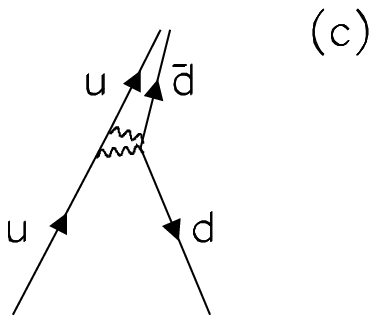,width=3cm}}
\vspace{0.5cm}
\caption{ Diagrams for $Z$-factors in the reaction
$\rho^+\to\gamma\pi^+$: a) $Z=e_d=-\frac 13$ and
b) $Z=e_u=\frac 23$.} \label{5}
\end{figure}

\begin{figure}[h]
%Fig. 7
\vspace{-0.5cm}
\centerline{\epsfig{file=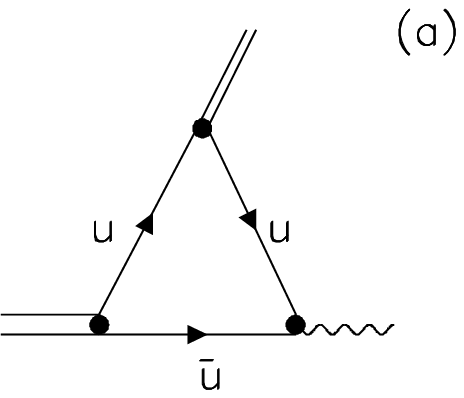,width=3cm}\hspace{1cm}
            \epsfig{file=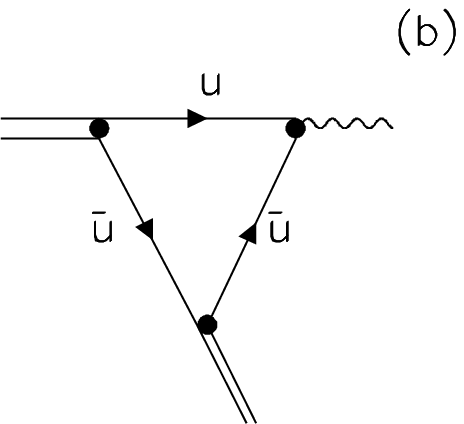,width=3cm}\hspace{1cm}
            \epsfig{file=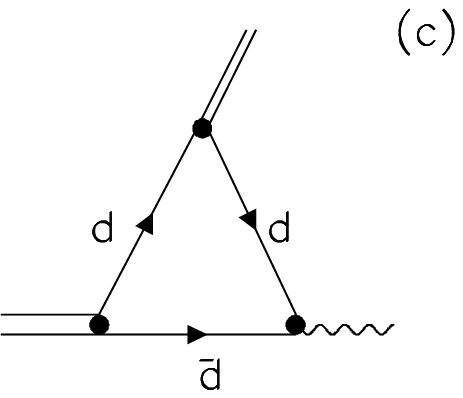,width=3cm}\hspace{1cm}
            \epsfig{file=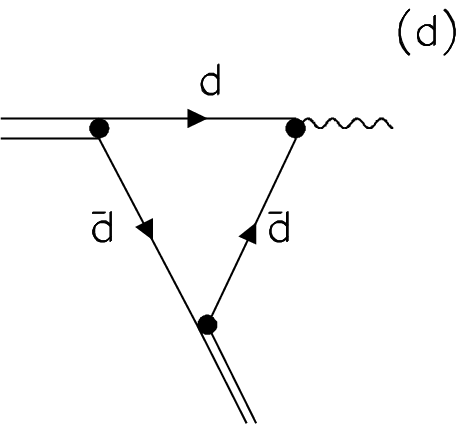,width=3cm}}
\vspace{0.5cm}
\caption{ Diagrams for $Z$-factors in the reactions $\rho^0\to\gamma\pi^0$
and $\omega^0\to\gamma\pi^0$:
a)  $Z(\rho^0\to\gamma\pi^0)=\frac 13$, $Z(\omega\to\gamma\pi^0)=\frac 13$;
b)$Z(\rho^0\to\gamma\pi^0)=\frac13$, $Z(\omega^0\to\gamma\pi^0)=\frac13$;
c)$Z(\rho^0\to\gamma\pi^0)=-\frac{1}{6}$, $Z(\omega^0\to\gamma\pi^0)=\frac16$;
d)$Z(\rho^0\to\gamma\pi^0)=-\frac{1}{6}$, $Z(\omega^0\to\gamma\pi^0)=\frac16$.
Recall that $\rho^0=\pi^0 = \frac{u\bar u -d\bar d}{\sqrt{2}}, \quad$ and
$\omega^0 = \frac{u\bar u + d\bar d}{\sqrt{2}}$. }\label{6}
\end{figure}

The charge factors for the pion emission in the decays
$\rho^+\to\gamma\pi^+$, $\rho^0\to\gamma \pi^0$, $\omega\to\gamma \pi^0$
(see Figs. \ref{5}, \ref{6})
 are equal to those for photon emission as follows:
\bea \label{dsi-17p}
&&Z^{\rho^+\to\gamma \pi^+}_{\bigtriangleup^\pi}=
Z^{\rho^+\to\gamma \pi^+}_{\bigtriangledown_\gamma}=e_d,
\qquad Z^{\rho^+\to\gamma \pi^+}_{\bigtriangledown_\pi}
= Z^{\rho^+\to\gamma \pi^+}_{\bigtriangleup^\gamma}=e_u\ , \nn \\
&&
Z^{\rho^0\to\gamma \pi^0}_{\bigtriangleup^\pi}=
Z^{\rho^0\to\gamma \pi^0}_{\bigtriangledown_\pi}=
Z^{\rho^0\to\gamma \pi^0}_{\bigtriangleup^\gamma}=
Z^{\rho^0\to\gamma \pi^0}_{\bigtriangledown_\gamma}=
\frac 12 (e_u+e_d), \nn \\
&&
Z^{\omega\to\gamma \pi^0}_{\bigtriangleup^\pi}=
Z^{\omega\to\gamma \pi^0}_{\bigtriangledown_\pi}=
Z^{\omega\to\gamma \pi^0}_{\bigtriangleup^\gamma}=
Z^{\omega\to\gamma \pi^0}_{\bigtriangledown_\gamma}=
\frac 12 (e_u-e_d).
\eea
In the calculation of $Z$-factors (\ref{dsi-17p}), we take into account that
pion emission by quark is a two-step process (see Fig. \ref{5}c ): the initial
quark (for example, in Fig. \ref{5}a) emits gluons (they have isospin $I_{gluons}=0$)
which produce quark--antiquark pairs, $u\bar u$ or $d\bar d$, with equal
amplitudes, and then we face the transition $u\bar d\to\pi^+ $. The block of
Fig. \ref{5}c is denoted  as a coupling $g_\pi$.

In the process of Fig. \ref{6}a, the gluons produce $u\bar u$ pair with the same
amplitude as in the previous case but then
we face the transiton $u\bar u\to\pi^0 $  resulting in the factor $1/\sqrt 2$
(recall that  $\pi^0=(u\bar u - d\bar d)/\sqrt 2$).
In the process of Fig. \ref{6}c, the $d\bar d$ pair is produced, and the
transiton $d\bar d\to\pi^0 $ gives the factor $- 1/\sqrt 2$ (for more
detailed presentation of the quark combinatorial rules see \cite{book3}
and references therein).

\subsubsection{Partial width}
In terms of the calculated form factors, the partial width reads:
\bea
\label{dsi-19p}
M_V\Gamma_{V\to\gamma \pi}&=&
\frac13\cdot \frac{\alpha}{4}\frac{M^2_V-M^2_\pi}{M^2_V}
\frac{\lambda(M^2_V,M^2_\pi,0)}{2}
\nn \\
&\times& \left[
 Z^{V\to\gamma \pi}_{\bigtriangleup^\gamma}
F^{V\to\gamma \pi}_{\bigtriangleup^\gamma}(0,M^2_\pi)
+ Z^{V\to\gamma \pi}_{\bigtriangleup^\pi}
F^{V\to\gamma \pi}_{\bigtriangleup^\pi}   (0,M^2_\pi)
\right .
 \nn \\
&&\left .+
 Z^{V\to\gamma \pi}_{\bigtriangledown_\gamma}
F^{V\to\gamma \pi}_{\bigtriangledown_\gamma} (0,M^2_\pi)
+ Z^{V\to\gamma \pi}_{\bigtriangledown_\pi}
F^{V\to\gamma \pi}_{\bigtriangledown_\pi}    (0,M^2_\pi)
\right]^2 .
\eea
Here, the factor $1/3$ is due to the averaging over initial vector meson
spin states, the term \\$\alpha/4 \ \cdot (M^2_V-M^2_\pi)/M^2_V$ is given
by the phase space integration, and $\lambda(M^2_V,M^2_\pi,0)/2=
(M^2_V-M^2_\pi)^2/2$ is
due to the spin factor (\ref{PgV-3}). The $Z$-factors are as follows:\\
$Z^{\rho^0\to\gamma \pi^0}_{\bigtriangleup^{\gamma}}=1/6$,
$Z^{\rho^0\to\gamma \pi^0}_{\bigtriangledown_{\gamma}}=1/6$,
$Z^{\rho^0\to\gamma \pi^0}_{\bigtriangleup^{\pi^0}}=1/6$,
$Z^{\rho^0\to\gamma \pi^0}_{\bigtriangledown_{\pi^0}}=1/6$,\\
$Z^{\omega\to\gamma \pi^0}_{\bigtriangleup^{\gamma}}=1/2$,
$Z^{\omega\to\gamma \pi^0}_{\bigtriangledown_{\gamma}}=1/2$,
$Z^{\omega\to\gamma \pi^0}_{\bigtriangleup^{\pi^0}}=1/2$,
$Z^{\omega\to\gamma \pi^0}_{\bigtriangledown_{\pi^0}}=1/2$.

\section{Results and discussion}

The fitting to the partial widths
$\Gamma^{(exp)}_{\rho^{\pm} \to\gamma\pi^{\pm}}=68\pm 30 $ keV,
$\Gamma^{(exp)}_{\rho^{0} \to\gamma\pi^0}=77\pm 28  $ keV,
$\Gamma^{(exp)}_{\omega \to\gamma\pi^0}=776\pm 45 $ keV leads to the following
values of the pion emission coupling:
\bea
\label{RD-1}
{\rm Solution \, I}  &:&\qquad \;\;\;  16.7 \pm 0.3\ ^{+0.1}_{-2.3}\ ,\nn \\
{\rm Solution \, II} &:&\qquad        -3.0  \pm 0.3\ ^{+2.1}_{-0.1}\ .
 \eea 
In Eq. (\ref{RD-1}), we have included systematical errors
($(+0.1/-2.3)$ for Solution I and $(+2.1/-0.1)$ for Solution II) which
are caused by the uncertainties of the fit of $q\bar q$ wave functions in the
spectral integral equation (see Section 1.2).

So, we have  regions of positive and negative $g_\pi$. However, one should
take into account that the sign of $g_\pi$ in (\ref{RD-1}) is rather
conventional: it depends on signs of wave functions of photon and
mesons involved into calculation. Because of that, being precise,
we should state that for $g_\pi$ we determine absolute values only,
see (\ref{g-pi}).

Solution I gives us the value of the of pion--nucleon coupling;
recall that it is determined
as a factor in the phenomenological Lagrangian:
$g_{\pi NN}\bigg(\bar \psi\, '_N (\vec\tau\vec\varphi_\pi)i\gamma_5\psi_N\bigg)$).
It is in agreement with the results for pion--nucleon scattering
$ g_{\pi NN}^2/4\pi\simeq 14 $ \cite{stoks,arndt,bugg-g}. Namely,
dealing with
pion--nucleon interaction in terms of the quark
 model, we use the Lagrangian:
\bea \label{g-pi1a}
g_{\pi qq}\bigg (\bar \psi\, '_q (\vec \tau \vec \varphi_\pi)\ i\gamma_5 \psi_q \bigg )
&=&
 \sqrt{2} g_{\pi qq}\, \varphi^+_{\pi^+}\bigg (
 \bar \psi\, '_d\ i\gamma_5 \psi_u   \bigg )+{\rm other\, terms} \nn\\
&=& g_{\pi }\, \varphi^+_{\pi^+}\bigg (
 \bar \psi\, '_d\ i\gamma_5 \psi_u
  \bigg )+{\rm other\, terms},  
\eea
that gives us  $\sqrt{2} g_{\pi qq}= g_{\pi }$.

In Appendix A, using SU(6)-symmetry for nucleons, we demonstrate
that $ g_{\pi NN}=(5/3)g_{\pi qq}$. So, in terms of  SU(6)-symmetry,
we have:
\be \label{g-pi2a}
 g_{\pi NN}=\frac{5}{3\sqrt{2}}g_{\pi}.
 \ee
 We see that Solution I, being in agreement with data
 \cite{stoks,arndt,bugg-g}, gives us
 \be \label{g-pi3}
g_{\pi NN}^2/(4\pi)= 22.2\pm 0.8^{+0.2}_{-5.0}.
 \ee

For Solution II, we have found
$ 0.03\le  g_{\pi NN}^2/(4\pi) \le  1$, that is far from the experimental
value.

\subsection{Predictions for excited vector states}

For $\rho^\pm(2S)$ , $\rho^0(2S)$ and $\omega(2S)$ mesons, we have found
the following partial widths (in keV units):
\bea
&&\Gamma (\rho_{2S}^\pm\to \gamma\pi)\simeq 10   - 130\, , \nn \\
&&\Gamma (\rho_{2S}^0\to \gamma\pi)\simeq 10   -130 \, , \nn \\
&&\Gamma (\omega_{2S}\to \gamma\pi)\simeq 60 - 1080\, .
\eea
The other wave functions of highly exited states have too large uncertainties
to provide us with reliable widths. This points to the necessity to carry out
mesurements of radiative processes with mesons in the region of large masses.

\subsubsection*{Acknowledgement}

We thank B.L. Birbrair for helpful remarks.
This paper was supported by the RFFI grant 07-02-01196-a.

\section*{Appendix A: Nucleon pion emission vertex \\ in the SU(6)
quark model}

Here we derive the relations between couplings in
phenomenological Lagrangian for pions and nucleons,
$g_{\pi NN}\bigg(\bar \psi\, '_N (\vec\tau\vec\varphi_\pi)i\gamma_5\psi_N\bigg)$,
and those for quarks,
$g_{\pi qq}\bigg(\bar \psi\, '_q (\vec\tau\vec\varphi_\pi)i\gamma_5\psi_q\bigg)$.
To be definite, we consider transitions $p^\uparrow \to\pi^+ + n^\downarrow$ and
$u^\uparrow \to\pi^+ + d^\downarrow$. We use the following SU(6) wave functions
(see, for example, Appendix D in Ref. \cite{book2}):
\bea  \label{A-B1}
\psi_p\equiv p^\uparrow \bigg(q(1)q(2)q(3)\bigg) =
\frac{\sqrt 2}{3}(u^\uparrow u^\uparrow d^\downarrow
+d^\downarrow u^\uparrow u^\uparrow +
u^\uparrow d^\downarrow u^\uparrow )\nn \\
-
\frac{1}{3\sqrt 2}(u^\uparrow u^\downarrow d^\uparrow
+ d^\uparrow u^\uparrow u^\downarrow
+u^\downarrow d^\uparrow u^\uparrow
+u^\downarrow u^\uparrow d^\uparrow   +d^\uparrow u^\downarrow u^\uparrow
+u^\uparrow d^\uparrow u^\downarrow ), \nn \\
\bar\psi_n\equiv n^\downarrow \bigg(q(1)q(2)q(3)\bigg) =
\frac{\sqrt 2}{3}(d^\downarrow d^\downarrow u^\uparrow
+u^\uparrow d^\downarrow d^\downarrow +
d^\downarrow u^\uparrow d^\downarrow ) \\
-
\frac{1}{3\sqrt 2}(d^\downarrow d^\uparrow u^\downarrow
+ u^\downarrow d^\downarrow d^\uparrow
+d^\uparrow u^\downarrow d^\downarrow
+d^\uparrow d^\downarrow u^\downarrow
+u^\downarrow d^\uparrow d^\downarrow
+d^\downarrow u^\downarrow d^\uparrow ) . \nn
\eea
Recall that for baryon quarks we use notation of the type
$d^\downarrow u^\downarrow d^\uparrow\equiv
d^\downarrow (1) u^\downarrow (2)d^\uparrow (3)$.

The isospin block reads:
\be \label{A-B2}
(\vec\tau\vec\varphi_\pi)=
\sqrt{2}\ \frac{\tau_{1}+i\tau_{2}}{2}\
\frac{\varphi^{(1)}_\pi-i \varphi^{(2)}_\pi}{\sqrt 2} +
\sqrt{2}\ \frac{\tau_{1}-i\tau_{2}}{2}\
\frac{\varphi^{(1)}_\pi+i \varphi^{(2)}_\pi}{\sqrt 2} +
\tau_{3}\varphi^{(3)}_\pi  .
\ee
Transition  $p^\uparrow \to\pi^+ + n^\downarrow$ is given by the following
terms in nucleon and quark spaces:
\bea  \label{A-B3}
&&
g_{\pi NN}\langle \pi^+ n^\downarrow|\,
\sqrt{2}\ \frac{\tau_{1}+i\tau_{2}}{2}\
\frac{\varphi^{(1)}_\pi-i \varphi^{(2)}_\pi}{\sqrt 2}\,
i\gamma_5|p^\uparrow \rangle \nn \\
&&=
g_{\pi qq}\langle \pi^+ n^\downarrow \bigg(q(1)q(2)q(3)\bigg)|\,
\sqrt{2} \nn \\
&&\times\sum\limits_{j=1,2,3}\frac{\tau_{1}(j)+i\tau_{2}(j)}{2}\
\frac{\varphi^{(1)}_\pi-i \varphi^{(2)}_\pi}{\sqrt 2}\,
i\gamma_5(j)\; |p^\uparrow \bigg(q(1)q(2)q(3)\bigg) \rangle \ ,
\eea
where $\tau_{1}(j)$, $\tau_{2}(j)$ and $\gamma_5(j)$ act on $q(j)$.

In the non-relativistic limit, which we use for nucleons and
 constituent quarks,
$$
    i\gamma_5
  \to  (-i)(\vec\sigma \vec q)
$$
and direct calculations give:
\be \label{A-B4}
g_{\pi NN}= g_{\pi qq}\cdot 3\cdot\frac 59\ .
\ee
To simplify the calculations which lead to (\ref{A-B4}),
one can fix the direction of photon momentum,
for example, $ \vec q =(q_x,0,0)$ and then use
$\bigg(\vec q\vec \sigma(j)\bigg) \to q_x\sigma_1(j)$.


\begin{thebibliography}{99} \fussy
\bibitem{raddecay}
A.V. Anisovich, V.V. Anisovich, V.N. Markov,
M.A. Matveev, V.A. Nikonov and A.V. Sarantsev, J. Phys. G: Nucl. Part.
Phys. {\bf 31}, 1537 (2005).

\bibitem{book3}
A.V. Anisovich, V.V. Anisovich, M.A. Matveev, V.A. Nikonov,
J. Nyiri, A.V. Sarantsev, {\it "Mesons and baryons: systematisation and
methods of analysis"}, World Scientific, Singapore, 2008.


\bibitem{BS}
A.V. Anisovich, V.V. Anisovich, V.N. Markov,  M.A. Matveev, and
A. V. Sarantsev,
 Yad. Fiz. {\bf 67}, 794 (2004)
[Phys. At. Nucl., {\bf 67}, 773  (2004)].
\bibitem{CM}
G.F. Chew  and S. Mandelstam,
Phys. Rev.  {\bf 119}, 467 (1960).

\bibitem{QQ}
 V.V. Anisovich, L.G. Dakhno,
M.A. Matveev, V.A. Nikonov and A.V. Sarantsev,
 Yad. Fiz. {\bf 70}, 68 (2007)
[Phys. At. Nucl., {\bf 70}, 63  (2007)],
hep-ph/0510410;\\
 Yad. Fiz. {\bf 70}, 392 (2007)
[Phys. At. Nucl., {\bf 70}, 364  (2007)],
hep-ph/0511005.
\bibitem{qq}
 V.V. Anisovich, L.G. Dakhno,
M.A. Matveev, V.A. Nikonov and A.V. Sarantsev,
 Yad. Fiz. {\bf 70}, 480 (2007)
[Phys. At. Nucl., {\bf 70}, 450  (2007)],
 hep-ph/0511109.
\bibitem{h-e} V.V. Anisovich, {\it "Partons and constituent quarks in soft processes"}
 Proc. of the XIV PNPI Winter School, p. 3, Leningrad, 1979;\\
V.V. Anisovich, M.N. Kobrinsky, J. Nyiri, Yu.M. Shabelski {\it "Quark model
and high energy collisions"}, World Scientific, Singapore, 1985.


\bibitem{gluon}
G. Parisi and R. Petronzio, Phys. Lett. B {\bf 94}, 51 (1980);\\
M. Consoli and J.H. Field, Phys. Rev. D {\bf 49}, 1293 (1994).

\bibitem{cornwell}
J.M. Cornwell and J. Papavassiliou, Phys.  Rev. D {\bf40}, 3474 (1989).

\bibitem{gerasyuta}
V.V. Anisovich, S.M.  Gerasyuta, and A.V. Sarantsev, Int. J. Mod. Phys.
A {\bf 6}, 2625 (1991).

\bibitem{g-lat}
D.B. Leinweber {\em et al.}, Phys. Rev. D {\bf 58}, 031501 (1998).

\bibitem{bethe}
E. Salpeter and H.A. Bethe, Phys. Rev. {\bf 84}, 1232 (1951);\\
E. Salpeter, Phys. Rev. {\bf 91}, 994 (1953).

\bibitem{Hulth}
G. Hulth and H. Snellman, Phys. Rev D {\bf 24}, 2978 (1981).

\bibitem{Godfrey}
S. Godfrey and N. Isgur, Phys. Rev. D{\bf 32}, 189 (1985).


\bibitem{Lucha}
W. Lucha, F. Sch\"oberl, and D. Gromes, Phys. Rep. {\bf 200}, 127
(1991).


\bibitem{petry-meson}
R. Ricken, M. Koll, D. Merten, B.C. Metsch, and H.R. Petry, Eur. Phys.
J. A {\bf 9}, 221 (2000).

\bibitem{G}
D. Ebert, R.N. Faustov, and V.O. Galkin, Phys. Rev. D {\bf 67}, 014027
(2003).

\bibitem{Linde}
J. Linde and H. Snellman, Nucl. Phys. A {\bf 619}, 346 (1997).

\bibitem{M}S.N. M\"unz, Nucl. Rhys. A {\bf 609}, 364 (1996).

\bibitem{Gupta}S.N. Gupta, S.F. Radford, and W.W. Repko, Phys. Rev.
 D {\bf 54}, 2075 (1996).

\bibitem{Sch}G.A. Schuler, F.A Berends, and R. van Gulik, Nucl. Rhys.
B {\bf 523}, 423 (1998).

\bibitem{Huang}H.-W. Huang, {\it et. al.} Phys. Rev.
D {\bf 54}, 2123 (1996); D {\bf 56}, 368 (1997).


\bibitem{deut}
V.V. Anisovich, M.N. Kobrinsky, D.I. Melikhov, and A.V. Sarantsev, Nucl.
Phys. A {\bf 544}, 747 (1992); \\
A.V. Anisovich and V.A. Sadovnikova, Yad. Fiz. {\bf 55}, 2657 (1992);
{\bf 57}, 75 (1994); Eur. Phys. J. A {\bf 2}, 199 (1998).

\bibitem{physrev}
V.V. Anisovich, D.I. Melikhov, and V.A. Nikonov, Phys. Rev.  D {\bf52},
5295 (1995).

\bibitem{epja} A.V. Anisovich, V.V. Anisovich, and V.A. Nikonov,
 Eur. Phys. J. A {\bf 12}, 103 (2001).

\bibitem{YFscalar} A.V. Anisovich, V.V. Anisovich, V.N. Markov,
and V.A. Nikonov,
Yad. Fiz. {\bf 65}, 523 (2002) [Phys. At. Nucl.
{\bf 65}, 497 (2002)].

\bibitem{YFtensor}
A.V. Anisovich, V.V. Anisovich, M.A. Matveev, and V.A. Nikonov,
Yad. Fiz. {\bf 66}, 946 (2003)
[Phys. At. Nucl.
{\bf 66}, 914 (2003)].

\bibitem{operator}
A.V. Anisovich, V.V. Anisovich, V.N. Markov,
M.A. Matveev and A.V. Sarantsev, J. Phys. G: Nucl. Part. Phys.
{\bf 28}, 15 (2002).


\bibitem{PNPI-RAL}
A.V. Anisovich, C.A. Baker, C.J. Batty {\em et al.},
Phys. Lett. B{\bf449}, 114 (1999); B{\bf452}, 173 (1999);
B {\bf452}, 180 (1999); B {\bf452}, 187 (1999);
B {\bf472}, 168 (2000);  B {\bf476}, 15 (2000);  B {\bf477},
 19 (2000); B {\bf491}, 40 (2000);  B {\bf491}, 47 (2000);
B {\bf496}, 145 (2000);
B {\bf507}, 23 (2001);
B {\bf508}, 6 (2001); B {\bf513}, 281 (2001); B {\bf517}, 261 (2001);
B {\bf 517}, 273 (2001); \\ Nucl. Phys. A {\bf651}, 253 (1999);
A {\bf662}, 319 (2000); A {\bf662}, 344 (2000).

\bibitem{Bar} D. Barberis {\em et al.} (WA 102 Collab.), Phys. Lett.
B~{\bf471}, 440 (2000).
\bibitem{LL} R.S. Longacre and S.J. Lindenbaum, Report BNL-72371-2004;
Phys. Rev. D {\bf 70}, 094041 (2004).
\bibitem{L3-KK}  V.A. Schegelsky, A.V. Sarantsev and V.A. Nikonov,
A.V. Anisovich, Eur. Phys. J. A {\bf 27}, 207 (2006).

\bibitem{L3-3pi}  V.A. Schegelsky, A.V. Sarantsev,
A.V. Anisovich and M.P. Levchenko, Eur. Phys. J. A {\bf 27}, 199 (2006).



%\bibitem{ufn04}V.V. Anisovich, UFN, {\bf174}, 49 (2004)
%[Physics-Uspekhi, {\bf47}, 45 (2004)].


\bibitem{bugg} D.V. Bugg, Phys. Rep. {\bf 397}, 257 (2004).


\bibitem{klempt-z}
E. Klempt and A. Zaitsev, Phys. Rept. {\bf 454}, 1 (2007).



\bibitem{syst}
A.V. Anisovich, V.V. Anisovich, and A.V.~Sarantsev, Phys. Rev.
D~{\bf62}, 051502(R) (2000).


\bibitem{g-qq}
A.V. Anisovich, V.V. Anisovich, L.G. Dakhno, V.A. Nikonov, and V.A.
Sarantsev,
Yad. Fiz. {\bf 68}, 1892 (2005) [Phys. Atom. Nucl. {\bf 68}, 1830
(2005)].

\bibitem{PR}
V.V. Anisovich, D.I. Melikhov, V.A. Nikonov,
Phys. Rev. D {\bf 55}, 2918 (1997).
\bibitem{stoks} V. Stoks, R. Timmermans, J.J. de Swart, Phys. Rev. C {\bf 47},
512 (1993).
\bibitem{arndt} R.A. Arndt, I.I. Strakovsky and R.L. Workman,
 Phys. Rev. C {\bf 50}, 2731 (1994);
 ArXiv:nucl-th/9506005.
\bibitem{bugg-g} D.V. Bugg, R. Marchleidt, {\it "$\pi NN$ coupling constant from
$NN$ elastic data between 210 800 MeV"}, preprint NUCL-TH-9404017 (1994).
\bibitem{EPJA}
A.V. Anisovich, V.V. Anisovich and V.A. Nikonov, Eur. Phys. J.
A {\bf 12}, 103 (2001).

\bibitem{YF}
A.V. Anisovich, V.V. Anisovich, M.A. Matveev and V.A. Nikonov,
Yad. Fiz. {\bf 66}, 946 (2003) [Phys. Atom. Nucl. {\bf 66}, 914 (2003)].

\bibitem{book2}
V.V. Anisovich, M.N. Kobrinsky, J. Nyiri, Yu.M. Shabelski {\it "Quark model
and high energy collisions"}, 2nd edition, World Scientific, Singapore, 2004.

\end{thebibliography}
 \end{document}